\documentstyle[preprint,aps,epsf,amsmath]{revtex} 
\begin{document} 
\tighten 
\draft 
\preprint{UG-31/99} 
\title{Nonperturbative infrared dynamics of three-dimensional QED with 
a four-fermion interaction} 
\author{Valery Gusynin$^{1}$, Anthony Hams$^{2}$, and Manuel Reenders$^{2}$} 
\address{$^1$
Bogolyubov Institute for Theoretical Physics,\\
03143 Kiev, Ukraine}
\address{$^{2}$ 
Institute for Theoretical Physics,\\ 
University of Groningen, 
9747 AG Groningen, The Netherlands} 
\date{\today} 
\maketitle 
\begin{abstract} 
A nonlinear Schwinger-Dyson (SD) equation for the gauge boson propagator 
of massless QED in one time and two spatial dimensions is studied. It is 
shown that the nonperturbative solution leads to a nontrivial 
renormalization-group infrared fixed point quantitatively close to the 
one found in the leading order of the $1/N$ expansion,
with $N$ the number of fermion flavors.
In the gauged Nambu--Jona-Lasinio (GNJL) model an equation for the Yukawa 
vertex is solved in an approximation given by the one-photon exchange and
an analytic expression is derived for the propagator of 
the scalar fermion-antifermion composites.
Subsequently,
the mass and width of the scalar composites near 
the phase transition line are calculated as functions of the four-fermion 
coupling $g$ and flavor number $N$. The possible relevance of these results 
for describing
particle-hole excitations, in particular antiferromagnetic correlations, 
observed in the underdoped cuprates, is briefly discussed.
\end{abstract} 
\section{Introduction} 

Quantum electrodynamics in $2+1$ dimensions (${\rm QED}_3$) has
attracted much interest over recent years. Its version with $N$
flavors of massless four-component Dirac fermions shares a number of
features, such as confinement and chiral symmetry breaking,
with four-dimensional quantum chromodynamics ($\rm QCD$).
The loop expansion of a massless theory suffers from severe infrared 
divergencies. However, in the $1/N$ expansion, the theory becomes infrared 
finite \cite{Appelquist1}, with the effective dimensionless coupling
\begin{equation}
\bar\alpha(p)=\frac{e^2}{p(1+\Pi(p))},\qquad \Pi(p)=\frac{e^2N}{8p},
\qquad p=\sqrt{p^2},
\label{effcoupling}
\end{equation}
giving rise to the renormalization-group $\beta$ function:
\begin{equation}
\beta_{\bar\alpha}(\bar\alpha)\equiv p\frac{d\bar\alpha(p)}{dp}
=-\bar\alpha(1-\frac{N}{8}\bar\alpha).
\end{equation}
In Eq.~(\ref{effcoupling}) $e$ is the dimensionful gauge coupling and
$\Pi(p)$ is the polarization operator. At large momenta 
($p\gg\alpha\equiv e^2N/8$) the effective coupling (\ref{effcoupling})
approaches zero (asymptotic freedom) while  for small momenta ($p\ll\alpha$)
it runs to the infrared (IR) fixed point $8/N$. Here, the dimensionful 
parameter $\alpha$ plays a role similar to the $\Lambda_{QCD}$ scale. 
Since ${\rm QED}_3$ is a super-renormalizable theory, the running of the 
coupling should be understood as a Wilsonian rather than Gell-Mann-Low type,
and it is not associated with ultraviolet divergencies.

By studying the Schwinger-Dyson (SD) equation for the fermion self-energy
in leading order of the $1/N$ expansion, it was found in 
Ref.~\cite{Appelquist2} that a phase transition occurs when the coupling 
at the IR fixed point exceeds 
some critical value ($8/N>\pi^2/4$). This means there exists a
critical number of fermions $N_{\rm cr}$ ($N_{\rm cr}=32/\pi^2\simeq3.24$) 
below which dynamical mass generation takes place and above which the fermions
remain massless. This is similar to what happens in quenched ${\rm
QED}_4 $ \cite{MNFK,SCQED4}, where the gauge coupling must
exceed a critical value for chiral symmetry breaking to occur.
The appearance of a dimensionless critical coupling can be traced to
the scale invariant behavior of both theories. The scale invariance of
${\rm QED}_3$ is associated with the IR fixed point, since, as is
evident from Eq.~(\ref{effcoupling}), in the limit $p\ll\alpha$ the
dimensional parameter $e$ drops out of the running coupling (as well
as from SD equations for the Green's functions). Related to this is the
fact that the chiral symmetry breaking phase transition in both theories
belongs to a special universality class called conformal phase
transition (CPT) introduced in Ref.~\cite{miya97}. It is characterized
by a scaling function having an essential singularity at the
transition point, and by abrupt change of the spectrum of light excitations
as the critical point is crossed (for details about the CPT in
${\rm QED}_3$ see Ref.~\cite{GMS}).

The presence of a critical $N_{\rm cr}$ in ${\rm QED}_3$ is intriguing
especially because of possible existence of an analogous critical
fermion number $N_f=N_{\rm cr}$ in $(3+1)$-dimensional SU$(N_c)$ gauge
theories, as is suggested by both analytical studies
\cite{Banks,apptewij,miya97} and lattice computer simulations 
\cite{Brown,Iwasaki}.
Also, a nontrivial IR fixed point in ${\rm QED}_3$ may be related to
nonperturbative dynamics in condensed matter, in particular, dynamics of
non-Fermi liquid behavior \cite{dorey,aitchison}.

The fact that the value of the IR fixed point determines the critical
$N_{\rm cr}$, below which the system is in the symmetry broken phase, 
and that this critical value is found to be of order $3$ provides
motivation for searches beyond the $1/N$ expansion. It is especially
important because there is still controversy concerning the existence
of finite $N_{\rm cr}$ in ${\rm QED}_3$; some authors argue that the
generation of a fermion mass occurs at all values of $N$ 
\cite{Pennington,Pisarski}
what might mean the absence of the IR fixed point for the running
coupling.\footnote{This would happen, for example, if one finds more
soft behavior of the polarization operator in the infrared, like
$\Pi(p)\sim(\alpha/p)^\gamma$, with $\gamma<1.$} 
Despite the arguments of Ref.~\cite{ma96} 
in addition to the fact that studies of $1/N^2$ corrections 
to the gap equation showed the increase of the 
critical value ($N_{\rm cr}=128/3\pi^2\simeq 4.32$) 
\cite{Nash,Aitchison97,Gusynin96}, 
the situation is far from being conclusive. 
What we need is some kind of
self-consistent equation for the running coupling which is to be solved
nonperturbatively.

In the present paper we study such a nonlinear equation for the
running coupling which is the analogue of the ladder approximation for
the fermion propagator.\footnote{Recently, in Ref.\cite{Nick} another
nonlinear equation for the running coupling was proposed in order to
study nontrivial infrared structure of the theory. However, their
definition of the running coupling deviates considerably from the
standard one used in present paper and we will not attempt to compare
both approaches.} Similar to the gap equation, the
kernel is taken in the $1/N$ approximation, where it is nothing else 
as the one-loop photon-photon scattering amplitude with zero momentum 
transfer. 
The equation obtained is obviously gauge invariant. We then
study our equation both analytically and numerically. 
We find that the
vacuum polarization operator, obtained as a nonperturbative solution
of the equation, has the same infrared asymptotics as the one-loop
expression: $\Pi(p)\simeq C\alpha/p$, $C\simeq 1+1/14N$. Thus a
nontrivial IR fixed point persists in the nonperturbative solution. 
Moreover, the correction to the one-loop result ($C=1$) is small even 
at $N=1$
due to smallness of the numerical coefficient before $1/N$, that
explains why the leading order in the $1/N$ expansion (the one-loop
approximation) for the vacuum polarization works so well.

Further we proceed to studying ${\rm QED}_3$ with additional
four-fermion interactions (the gauged Nambu-Jona-Lasinio (GNJL) model).
Such kind of models are considered to be effective theories at
long distances in planar condensed matter physics, in particular, for
high temperature superconductivity \cite{high-T_c}. It is well known
that in the improved ladder approximation (with the photon propagator 
including fermion one-loop effects) this model has a nontrivial phase 
structure \cite{caclwa91a} in the coupling constant plane $(1/N,g)$, where
$g=2G\Lambda/\pi^2$ is the dimensionless four-fermion coupling
($\Lambda$ is the ultraviolet cutoff). The critical line is
\begin{equation}
g_c\left(1/N\right)=\frac{1}{4}\left(1+\sqrt{1-\frac{N_{\rm cr}}
{N}}\right)^2,\qquad N>N_{\rm cr},
\label{critline}
\end{equation}
at $g>1/4$, and $1/N=1/N_{\rm cr}$ at  $g<1/4$. Above this line the gap
equation for the fermion self-energy $\Sigma(p)$ has a nontrivial
solution. Thus the chiral symmetry is dynamically broken, which
implies the existence of a nonzero vacuum condensate
$\langle\bar\psi\psi\rangle$.
One end point $(1/N=0,g=1)$ of the critical line corresponds
to the ordinary NJL model (in $2+1$ dimensions ), while the other one
$(1/N=1/N_{\rm cr},g=0)$ corresponds to pure ${\rm QED}_3$.

A nice feature of this model is that it is renormalizable in the $1/N$
expansion \cite{rowapa89a} leading to an interacting continuum
($\Lambda\to\infty$) theory near a critical scaling region
(critical curve) separating a chiral symmetric phase ($\chi S$) and
a spontaneous chiral symmetry broken phase ($S\chi SB$).
The spectrum of such a theory contains pseudoscalar ($\pi$) and
scalar ($\sigma$) bound states which become light in the vicinity of 
the critical line.
Since the phase transition is second order along the $g>1/4$  part
Eq.~(\ref{critline}) of the critical curve, scalar and pseudoscalar
resonances are to be produced on the symmetric side of the curve,
whose masses approach zero as the critical curve is approached
\cite{aptewij91,gure98}.
The part of the critical curve with $g<1/4, 1/N=1/N_{\rm cr}$ is
rather special and is related to the CPT in pure ${\rm QED}_3$ (we
shall discuss it more in the main text).

In this work we study scalar composites ($\sigma$ and $\pi$ bosons)
which are resonances in the symmetric phase of the $2+1$-dimensional 
GNJL model. The $\pi$ boson can be viewed as a Goldstone boson precursor
mode that comes down in energy as the transition is approached.
Our study is motivated partially by possible relation of these resonances 
to spin excitations observed in neutron scattering experiments in underdoped 
high-$T_c$ superconductors \cite{kile99}.
We calculate their masses and widths as a function of the four-fermion
coupling $g$ and therefore mass and width's dependence on the doping
concentration (since in certain low-energy effective models based on
spin-charge separation, the coupling $g$ would depend on the doping, 
{\em e.g.}, Ref.~\cite{Mavromatos}).

The plan of the present paper is as follows. In Sec.~\ref{sec_QED3}
we derive a nonlinear equation for the effective running coupling in
pure ${\rm QED}_3$ which is then solved 
both analytically and numerically 
to establish the existence of a nontrivial IR fixed
point. In Sec.~\ref{sec_GNJL2+1} after introducing the GNJL model in
$2+1$ dimensions we solve the equation for the Yukawa
vertex with nonzero boson momentum. In Sec.~IV we obtain an analytical
expression for the boson propagator valid along
the entire critical line and analyze its behavior in different
asymptotical regimes. The analysis of the scalar composites near the
critical line (\ref{critline}) is given in Sec.~V. 
We present our summary in Sec.~\ref{conclusion}. 
In Appendix~\ref{app:box} we compute the one-loop photon-photon 
scattering amplitude with zero transferred momentum
and list some useful angular integrals.
In Appendix~\ref{app_nlg}, an expression for the
nonlocal gauge for the ladder (bare vertex) approximation is derived.
Finally, in Appendix~\ref{app_twochan},
we present some details of the approximation \cite{gure98}
which is used to solve the equation for the Yukawa vertex.
\section{The equation for the running coupling in 
{\protect\boldmath${\rm QED}_3$}}\label{sec_QED3}
The Lagrangian density of massless ${\rm QED}_3$ in a general
covariant gauge is given by
\begin{equation}
{\cal L}=-\frac{1}{4}F^2_{\mu\nu}+\frac{1}{2a}(\partial_\mu A^\mu)^2+
\bar\psi_i i\gamma^\mu D_\mu\psi_i,
\end{equation}
where $D_\mu=\partial_\mu-ieA_\mu$ is the covariant derivative.
In a parity invariant formulation we consider $N$ flavors of fermions 
($i=1,\dots N$) described by four-component spinors.
The three $4\times 4$ $\gamma$-matrices are taken to be
\begin{eqnarray}
&&
\gamma^0=\left(\begin{array}{cc} \sigma_3 & 0 \\ 0 & -\sigma_3\end{array}
         \right),\quad
\gamma^1=\left(\begin{array}{cc} i\sigma_1 & 0 \\ 0 & -i\sigma_1\end{array}
         \right),\quad
\gamma^2=\left(\begin{array}{cc} i\sigma_2 & 0 \\ 0 & -i\sigma_2\end{array}
         \right),
\end{eqnarray}
with $\sigma_i$ the usual Pauli matrices. There are two matrices
\begin{eqnarray}
\gamma^3=i\left(\begin{array}{cc} 0 & 1 \\ 1 & 0\end{array}
         \right),\quad
\gamma^5=\gamma_5
=-i\left(\begin{array}{cc} 0 & 1 \\ -1 & 0\end{array}
         \right),
\end{eqnarray}
that anticommute with $\gamma^0$, $\gamma^1$ and $\gamma^2$. Therefore for
each four-component spinor, there is a global U$(2)$ symmetry with
generators $I$, $\frac{1}{i}\gamma^3$, $\gamma^5$, and 
$\frac{1}{2}[\gamma^3,\gamma^5]$, and the full symmetry is then U$(2N)$.
In what follows we shall restrict ourselves to the symmetric phase of
the model, {\em i.e.}, massless fermions.

\begin{figure}
\begin{center}
\epsfxsize=9cm
\epsffile{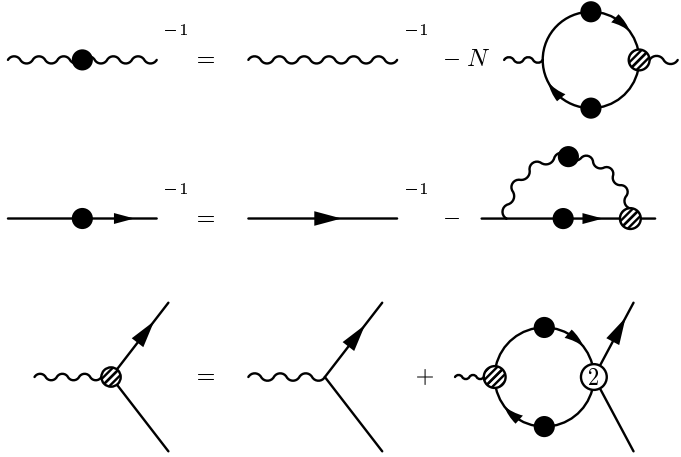}
\end{center}
\caption{Exact SD equations for the gauge boson propagator $D_{\mu\nu}$,
the fermion propagator $S$, and the vertex $\Gamma^\mu$.}
\label{fig:exact}
\end{figure}

The exact SD equations are given in Fig.~\ref{fig:exact}. For clarity
we have extracted the explicit factors of $N$ coming from the one-fermion 
loop.
Since in pure ${\rm QED}_3$ we have only one dimensionful parameter $e$, 
this enables us to choose our scale such that $N e^2$ remains fixed. 
This means that every photon
propagator (times $e^2$) contributes one factor of $1/N$.

\begin{figure}
\epsfxsize=7cm
\begin{center}
\epsffile{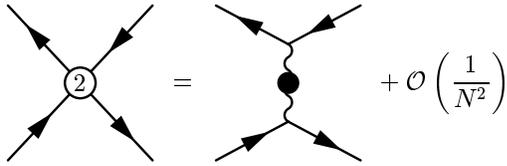}
\end{center}
\caption{The $1/N$ expansion of the two-fermion, one-photon irreducible
fermion-fermion scattering kernel.}
\label{fig:kern}
\end{figure}

\begin{figure}
\begin{center}
\epsfxsize=12cm
\epsffile{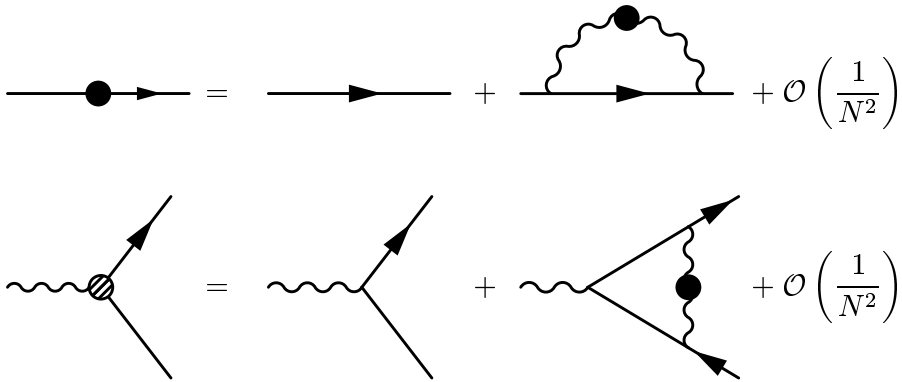}
\end{center}
\caption{SD equation for fermion propagator and vertex up to order $1/N^2$.}
\label{fig:trunc}
\end{figure}

\begin{figure}
\epsfxsize=14cm
$
\begin{array}{c}
\epsffile{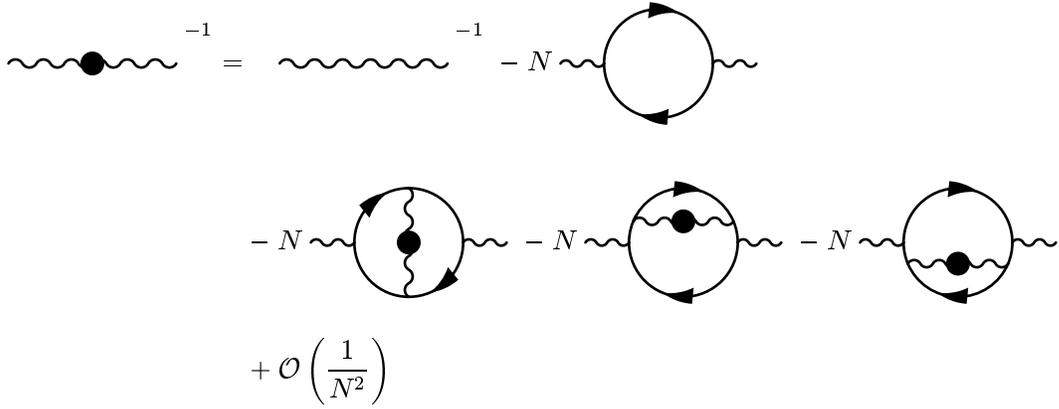}
\end{array}
$
\caption{Closed
SD equation for the gauge boson propagator in next-to-leading
$1/N$ expansion.}
\label{fig:expansion}
\end{figure}

To make a $1/N$ expansion of Fig.~\ref{fig:exact}, we first need to expand 
the two-fermion, one-photon irreducible fermion-fermion
scattering kernel, see Fig.~\ref{fig:kern}. We can convince ourselves that
Fig.~\ref{fig:kern} is indeed the right expansion, since the only corrections
of order one are fermion loops and they are already included in the full 
photon propagator.
Inserting this expansion into the SD equation for the vertex, we obtain 
a closed set of integral equations. 
The nice feature of this truncated system of SD equations is that it 
satisfies the Ward-Takahashi (WT) identities for the vertex as well as for the
vacuum polarization \cite{Gusynin96}.
However, finding an analytic solution seems to be a formidable task 
and further approximations are required. For such an approximation,
we simplify the fermion and the vertex SD equations by keeping only 
the lowest order terms in the $1/N$ expansion (see Fig.~\ref{fig:trunc}).
Then, inserting this into the SD equation for the photon 
propagator we obtain the equation shown in Fig.~\ref{fig:expansion}
which still satisfies the WT identity.
After that equation has been solved, the fermion propagator and the
vertex can be evaluated explicitly through right-hand sides of
Fig.~\ref{fig:trunc}.

We could solve the obtained photon propagator equation
by further iteration, with the one-loop fermion correction included 
at the initial step to obtain a perturbative $1/N$ expansion.
Instead we choose to solve the nonlinear integral equation given by
Fig.~\ref{fig:expansion} as it is.
In this way we might get a hint of any nonanalytic behavior in $1/N$ 
which would be lost otherwise.
At first glance, this way of solving a truncated system of SD equations
ignores possible nonanalyticity in $1/N$ coming from the fermion 
wave function and the vertex (related, for example, to the power-law
behavior due to an anomalous dimension). Note, however, that the fermion
propagator is a gauge dependent quantity, thus possible 
power-law behavior of the fermion wave function must cancel the 
corresponding behavior coming from the longitudinal part of the vertex
(recall that we consider in this paper the symmetric (massless) phase only).
Therefore the only nonanalyticity we have neglected is the one which might
be present in the transverse part of the vertex beyond order $1/N$.
Neglecting possible nonanalyticity in the 
transverse vertex means that we are seeking for nonanaliticity 
originating from the nonlinear equation for the photon propagator only.
To some extent, the considered approximation is similar to solving
the SD equation for the fermion mass function in the ladder 
approximation, where the photon propagator is taken in the leading 
$1/N$ order \cite{Appelquist1}.

In $2+1$ dimensions, the SD equation for the photon propagator reads
\begin{equation}
D^{-1}_{\mu\nu}(p)=D^{-1}_{0\,\mu\nu}(p)+\Pi_{\mu\nu}(p),
\end{equation}
with $D_{0\,\mu\nu}$ the bare photon propagator, and
where $\Pi_{\mu\nu}$ is the vacuum polarization tensor
\begin{eqnarray}
\Pi^{\mu\nu}(p)=i Ne^2
\int_M \frac{d^3r}{(2\pi)^3}\,
{\rm Tr}\left[\gamma^{\mu} S(r+p)\Gamma^{\nu}(r+p,r) S(r)\right].
\label{SDvacpolgb}
\end{eqnarray} 
Because of the gauge symmetry the vacuum polarization tensor is transverse:
\begin{eqnarray}
\Pi^{\mu\nu}(p)=(-g^{\mu\nu}p^{2}+p^{\mu}p^{\nu})\Pi(p), 
\label{eq:vacdef}
\end{eqnarray}
therefore, for the full photon propagator in a general covariant
gauge, we have
\begin{eqnarray} \label{eq:photonpropagator}
D_{\mu\nu}(p)=\left(-g_{\mu\nu}+\frac{p_\mu p_\nu}{p^2} \right)
\frac{1}{p^2}\frac{1}{\left[1+\Pi(p)\right]}-a\frac{p_\mu p_\nu}{p^4}.
\end{eqnarray}
Moreover, one can write
\begin{eqnarray}
\Pi(p)= -\frac{1}{2p^2}
\left( g_{\mu\nu}-c_1\frac{p_{\mu} p_{\nu}}{p^2} \right)
\Pi^{\mu\nu}(p),
\label{projection}
\end{eqnarray}
where the constant $c_1$ can be chosen arbitrarily.

\begin{figure}
$
\begin{array}{c}
\epsffile{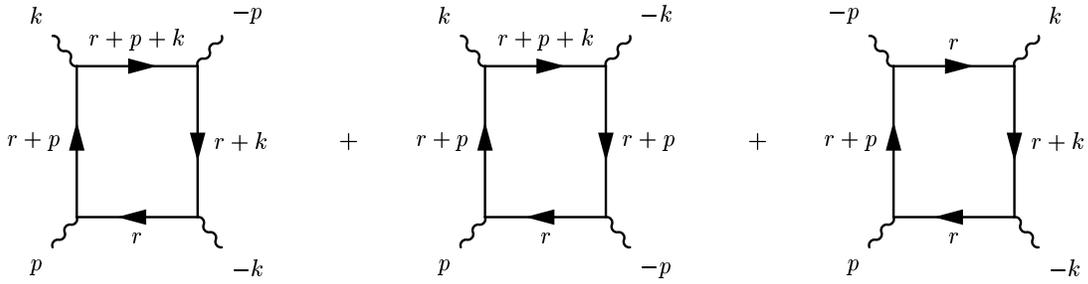}
\end{array}
$
\bigskip
\caption{The box diagram $B^{\mu\rho\nu\sigma}$.}
\label{fig:boxes}
\end{figure}

The vacuum polarization $\Pi(p)$ governs the running of the dimensionless
gauge coupling.
Now we study the integral equation based on Fig.~\ref{fig:expansion},
this gives
\begin{eqnarray}
\Pi^{\mu\nu}(p) = \Pi_1^{\mu\nu}(p) +\Pi_2^{\mu\nu}(p)
+ {\cal O}\left(1/N\right),
\label{eq:pitensor}
\end{eqnarray}
where $\Pi_1^{\mu\nu}(p)$ is the one-loop vacuum polarization,
\begin{eqnarray}
\Pi_1^{\mu\nu}(p)=i Ne^2
\int_M \frac{d^3r}{(2\pi)^3}\,
{\rm Tr}\left[\gamma^{\mu} S_0(r+p)\gamma^{\nu} S_0(r)\right],
\label{SDvacpol1loop}
\end{eqnarray}
with $S_0(p)$ the bare fermion propagator, $S_0(p)=1/\hat p$,
and
\begin{eqnarray}
\Pi_2^{\mu\nu}(p)=
 iNe^4 \int_M \frac{d^3k}{(2\pi)^3}\,
D_{\rho\sigma}(k) B^{\mu\rho\nu\sigma}(p,k), \label{pitens2}
\end{eqnarray}
where $B^{\mu\rho\nu\sigma}(p,k)$
is the one-loop ``photon-photon'' scattering amplitude, with zero momentum
transfer, {\em i.e.},
\begin{eqnarray}
B^{\mu\rho\nu\sigma}(p,k)&=&i\int_M\frac{d^3r}{(2\pi)^3}\,
{\rm Tr}\left[\gamma^{\mu} S_0(r+p)\gamma^{\rho}
S_0(r+p+k) \gamma^{\nu} S_0(r+k)\gamma^{\sigma} S_0(r)\right]
\nonumber\\
&+&i\int_M\frac{d^3r}{(2\pi)^3}\,
{\rm Tr}\left[\gamma^{\mu} S_0(r+p)
\gamma^{\rho} S_0(r+p+k)\gamma^{\sigma}
S_0(r+p) \gamma^{\nu} S_0(r)\right]\nonumber\\
&+&i\int_M\frac{d^3r}{(2\pi)^3}\,
{\rm Tr}\left[\gamma^{\mu} S_0(r+p)\gamma^{\nu} S_0(r)\gamma^{\rho}
S_0(r+k) \gamma^{\sigma} S_0(r)\right]. \label{boxeq1}
\end{eqnarray}
A graphical representation of the ``box'' diagram (\ref{boxeq1})
in terms of Feynman diagrams is given in Fig.~\ref{fig:boxes}. 

For the scattering amplitude $B^{\mu\rho\nu\sigma}$ there exists a 
Ward-Takahashi identity \cite{kane50}, 
which states the transversality of the
amplitude with respect to external photon momenta,
\begin{eqnarray}
p^{\mu} \,B_{\mu\nu\rho\sigma} (p,k,q,r) = 0,
\qquad
k^{\nu} \,B_{\mu\nu\rho\sigma} (p,k,q,r) = 0, \qquad
\mbox{etc.} \label{boxwti}
\end{eqnarray}

The vacuum polarization tensor has a superficial linearly divergent
part, which can be removed by a proper gauge-invariant regularization. 
However, since the divergent part is proportional to
$g_{\mu\nu}$ we can project out the finite vacuum polarization by
contracting $\Pi_{\mu\nu}(p)$ with the projector

\begin{eqnarray}
P_{\mu\nu}(p)=\left(g_{\mu\nu}-3\frac{ p_\mu p_\nu}{p^2}\right), 
\label{gmnkill}
\end{eqnarray}
{\em i.e.}, we choose the constant $c_1$ in (\ref{projection}) to be $c_1=3$.
This approach was used in Refs.~\cite{burden92} and \cite{ma96}.
In this way, we obtain
\begin{eqnarray}
\Pi(p)=\Pi_1(p) +\Pi_2(p)+ {\cal O}(1/N), \label{eq:piscalar}
\end{eqnarray}
with
\begin{eqnarray}
\Pi_1(p)&=&-\frac{4iNe^2}{p^2}\int_M\frac{d^3k}{(2\pi)^3}\,
\left[\frac{k^2-2k\cdot p-3(k\cdot p)^2/p^2}{k^2 (k+p)^2}\right],\\
\Pi_2(p)&=&\frac{iNe^4}{2p^2} \int_M\frac{d^3k}{(2\pi)^3}\,
\frac{B(p^2,k^2,p\cdot k)}{k^2\left[1+\Pi(k)\right]},
\end{eqnarray}
where
\begin{eqnarray}
B(p^2,k^2,p\cdot k)=
g_{\mu\nu} g_{\rho\sigma}B^{\mu\rho\nu\sigma}(p,k). \label{eq:bdef}
\end{eqnarray}

In Euclidean formulation the above expressions can be written as
\begin{eqnarray}
\Pi_1(p)&=&\frac{2Ne^2}{\pi^2 p}
\int_0^\infty dk\int\frac{d\Omega}{4\pi}\,
\left[\frac{k^2-2k\cdot p-3(k\cdot p)^2/p^2}{p(k+p)^2}\right]=\frac{Ne^2}{8p},
\label{1loopvac}
\\
\Pi_2(p)&=&-\frac{Ne^4}{4 \pi^2 p}
\int_0^\infty dk \,\frac{K(p,k)}{p\left[1+\Pi(k)\right]} \label{eq:pi2},
\end{eqnarray}
where
\begin{eqnarray}
K(p,k)\equiv \int\frac{d\Omega}{4\pi}\,B(-p^2,-k^2,-p\cdot k).\label{Kkerndef}
\end{eqnarray}

From Figs.~\ref{fig:expansion} and \ref{fig:boxes}, 
one can see that the first term in Eq.~(\ref{boxeq1}) corresponds to a vertex 
correction and the last two terms are fermion self-energy corrections.
The sum of these diagrams has symmetries which provide a consistency
check on the final result.
From the graphical representation it is obvious that the quantity 
$B(p^2,k^2,p\cdot k)$ should be invariant under $p \leftrightarrow k$
and under $p \rightarrow -p$ or $p\cdot k \rightarrow -p\cdot k$.

A detailed computation of the ``box'' function $B$ is presented in 
Appendix~\ref{app:box}, and the final expression for $B$ is given by 
Eq.~(\ref{boxBexpr1}).
One can verify that Eq.~(\ref{boxBexpr1}) has the symmetries we mentioned
above. Finally, we perform the angular integration to obtain $K(p,k)$,
\begin{eqnarray}
K(p,k)&=&
\,{\cal P}\int\frac{d\Omega}{4\pi}\,
\Biggr[\frac{1}{k}+\frac{1}{p}
+\frac{k\cdot p}{2kp|k-p|}
-\frac{2k^2+kp+2p^2}{kp|k-p|}
+\frac{2k^4+5k^2p^2+2p^4}{2(k\cdot p)kp|k-p|}
\Biggr]
\nonumber\\
&=& \frac{1}{k}+\frac{1}{p}
+\frac{kp}{6\,\mbox{max}(k^3, p^3)}
-\frac{2k^2+kp+2p^2}{kp\,\mbox{max}(k,p)}
\nonumber\\
&&
+\frac{2k^4+5k^2p^2+2p^4}{2k^2p^2 \sqrt{p^2+k^2}}
\sinh^{-1} \frac{\min(p,k)}{\max(p,k)}
\label{eq:k}
\end{eqnarray}
where we have made use of the integrals given in Appendix~\ref{app:box}.

Thus, we arrive at the following nonlinear equation for the vacuum
polarization:
\begin{equation}
\Pi(p)=\frac{Ne^2}{8p}-\frac{Ne^4}{4\pi^2 p}
\int_0^\infty dk \,\frac{K(p,k)}{p\left[1+\Pi(k)\right]}.
\label{vacpoleq}
\end{equation}
Apparently, this equation is gauge invariant.
We can rewrite it also as the equation for the running coupling 
$\bar\alpha(p)$ which must be
self-consistently determined from it:
\begin{equation}
\bar\alpha^{-1}(p)=\bar\alpha^{-1}_1(p)-\frac{N}{4\pi^2p}\int\limits_0^\infty
dk\,k K(p,k)\bar\alpha(k),
\label{effcoupleq}
\end{equation}
where $\bar \alpha_1(p)$ is the one-loop running coupling 
(see Eq.~(\ref{effcoupling})).

Equation~(\ref{effcoupleq}) is the simplest nonlinear equation for the
running coupling (or the photon propagator) which is derived at the
lowest order in the $1/N$ truncation of the SD equations. In fact, it should
be considered as an analogue of the ladder approximation for the
fermion propagator. The effects of a constant fermion mass can be incorporated
at one's wish by computing the box diagrams with massive fermions. This
would allow one to study the coupled system of the SDE for the fermion
self-energy and photon polarization operator along the lines of
Ref.~\cite{Gusynin96}.
However, this is beyond the scope of the present paper and we shall
leave aside this issue.\footnote{
A coupled system of SD equations for the vacuum polarization 
and the fermion renormalization wave function was studied 
in Ref.~\cite{ma96} using an Ansatz for the full vertex satisfying 
the Ward-Takahashi (WT) identity. 
Though such an approach reproduces standard value for the
critical $N_{\rm cr}\approx 3.3$, 
it does not permit us to identify the Ansatz with 
a class of Feynman diagrams.} 

Now we proceed by solving Eqs.~(\ref{vacpoleq}) and (\ref{effcoupleq}) 
both analytically and numerically. 
Approximating, as usual, the expression
(\ref{eq:k}) for the kernel by its asymptotics at $p \gg k$ and $p \ll k$
\begin{equation}
K(p,k) \simeq -\frac{2}{15} \frac{p^3 k^3}{\max(p^7,k^7)},
\end{equation}
one can reduce the integral Eqs.~(\ref{vacpoleq}) and
(\ref{effcoupleq}) to differential ones in order to study the 
asymptotical behavior of $\Pi(p)$ and $\bar\alpha(p)$ in the ultraviolet 
and infrared regions. However, in the present case we can find 
corresponding asymptotics directly from the integral equations.

First of all, we can immediately see that the solution of
Eq.~(\ref{effcoupleq}) for the running coupling possesses a
nontrivial IR fixed point. Indeed, by making a change of variables, $k\to
kp$, in the integral and assuming that $\bar\alpha(0)\neq 0$ we come to the
quadratic equation for $\bar\alpha(0)$:
\begin{equation}
\bar\alpha^{-1}(0)=\bar\alpha^{-1}_1(0)-\frac{N}{4\pi^2}\int\limits_0^\infty
dk\,k K(1,k) \bar\alpha(0),
\end{equation}
where we have made use of the fact that $p K(p,kp)=K(1,k)$, 
see Eq.~(\ref{eq:k}).
The last integral can be evaluated exactly 
(see Appendix~\ref{app:box}), and we obtain
\begin{eqnarray}
\bar\alpha(0)=\frac{8}{NC},
\quad C=\frac{1}{2}+\frac{1}{2}\sqrt{1+\frac{4}{N}
\left(\frac{184}{9\pi^2}-2\right)}\simeq
1+\frac{1}{N}\left(\frac{184}{9\pi^2}-2\right)\simeq
1+\frac{1}{14.0\, N}.
\label{Ceq}
\end{eqnarray}
This result illustrates that the $1/N$ expansion is reliable even
for a rather low number of flavors, {\em e.g.} $N=2$, because of the
smallness of the numerical coefficient in front of the $1/N$ term.

\begin{figure}
\epsfxsize=11cm
\epsffile{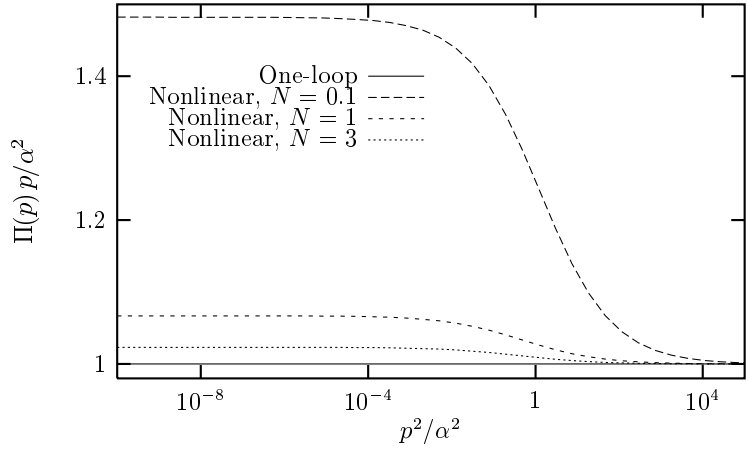}
\caption{Numerical solutions of Eq.~(\ref{vacpoleq}).}
\label{fig:numsol}
\end{figure}

The next term in the expansion of $\bar\alpha(p)$ at small $p$ can also be
calculated exactly, as well as its asymptotics at large momenta but 
we focus on finding the asymptotics of the vacuum polarization operator
itself. For it we seek a power solution ($\sim(p/\alpha)^\gamma$) in
both asymptotic regions, ($p\ll\alpha$) and ($p\gg\alpha$). We find
that the power exponent can only be $\gamma=-1$ in both cases. Thus 
we get
\begin{eqnarray}
\Pi(p)&=& C\frac{\alpha}{ p},\quad \mbox{for} \quad p \ll \alpha, 
\label{eq:IRpi}\\
\Pi(p)&=& \frac{\alpha}{p}, \quad \mbox{for} \quad p \gg \alpha 
\label{eq:UVpi}
\end{eqnarray}
with the constant $C$ defined in Eq.~(\ref{Ceq}) (we recall that 
$\alpha=e^2N/8$). Hence for the running coupling we have
\begin{eqnarray}
\bar\alpha(p)&=&\frac{e^2}{p(1+C\alpha/p)}\approx \frac{8}{CN},
\quad p\ll\alpha,\\
\bar\alpha(p)&=&\frac{e^2}{p(1+\alpha/p)},\quad p\gg\alpha.
\end{eqnarray}
The numerical solution of Eq.~(\ref{vacpoleq}) 
is presented in Fig.~\ref{fig:numsol}.
From this figure it is clear that the IR behavior ({\em i.e.}, $p\ll \alpha$) 
of $p \Pi(p)$ is indeed constant and in agreement with the 
analytic analysis.  

For studying effects like symmetry breaking and dynamical mass
generation, it is sufficient to consider only momenta less than $\alpha$. 
Therefore for the remainder of this article we will just
use Eq.~(\ref{eq:IRpi}) and treat $\alpha$ as the ultraviolet cutoff for
nonperturbative dynamics.
This allows us to write the gauge boson propagator as
(in Euclidean formulation)
\begin{equation}
e^2 D_{\mu\nu}(p)=\left(-g_{\mu\nu}+(1-\xi(p))\frac{p_\mu p_\nu}{p^2}
\right)\frac{\bar\alpha(0)}{p},\qquad \bar\alpha(0)=\frac{8}{NC}, 
\label{IRphot}
\end{equation}
for $p=\sqrt{p^2}\leq \alpha$, with $C$ given by Eq.~(\ref{Ceq}), and
where $\xi(p)$ parameterizes a nonlocal gauge fixing function 
(see Appendix~\ref{app_nlg}). This form of the photon propagator
will be used in the next section.

The gauge boson propagator of Eq.~(\ref{IRphot}) gives rise to a 
Coulomb potential instead of a logarithmically confining potential.
The dimensionless coupling $\bar\alpha_0\equiv\bar\alpha(0)$ should 
now be interpreted as the coupling parameter of a perfectly marginal
(or conformal invariant) interaction: $\beta(\bar\alpha_0)=0$.

\section{{\protect\boldmath ${\rm QED}_3$} plus Four-fermion 
interactions}\label{sec_GNJL2+1}

The gauged NJL model with $N$ fermion flavors is described by the 
Lagrangian
\begin{equation}
{\cal L}_{\rm GNJL}=
-\frac{1}{4} F^2_{\mu\nu}+\bar \psi (i\gamma^{\mu}D_{\mu}-m_0)\psi
+\frac{G}{2N}[(\bar\psi\psi)^2+(\bar\psi i\gamma_5\psi)^2],
\label{lag1}
\end{equation}
where  $D_{\mu}=\partial_{\mu}-ieA_{\mu}$ is the covariant derivative,
and the last term is a chirally invariant four-fermion interaction
with $G$ the corresponding Fermi coupling constant. In the absence
of a fermion mass term $m_0$ which breaks the chiral symmetry
explicitly, the Lagrangian (\ref{lag1}) possesses a U$(1)$ gauge
symmetry and a global U$_L(1)\times $U$_R(1)$ chiral symmetry. For
the four-fermion coupling we introduce the dimensionless coupling
constant $g=2G\Lambda/\pi^2$,
and we consider the dimensionful gauge coupling $e^2$ as the UV cutoff
(more precisely, $\alpha\simeq\Lambda$).

A parity invariant bare mass term $m_0\bar\psi\psi$ as well as a dynamically
generated fermion mass breaks the
global symmetry down to U$_{L+R}(1)$.  Further we study mainly the chiral
symmetric case with $m_0=0$. By introducing the auxiliary scalar
fields $\sigma$ and $\pi$, the Lagrangian (\ref{lag1}) can be
rewritten as
\begin{eqnarray}
{\cal L}_2=-\frac{1}{4} F_{\mu\nu}F^{\mu\nu}+\bar\psi
i\gamma^{\mu}D_{\mu}\psi-\bar\psi(\sigma+i\gamma_5\pi)\psi
-\frac{N}{2G}\left(\sigma^2+\pi^2\right),
\label{lag2}
\end{eqnarray}
where $\sigma=-(G/N)\bar\psi_i\psi_i$,
$\pi=-(G/N)\bar\psi_i\gamma_5\psi_i$.

\begin{figure}
\epsfxsize=9cm
\epsffile[30 400 360 540]{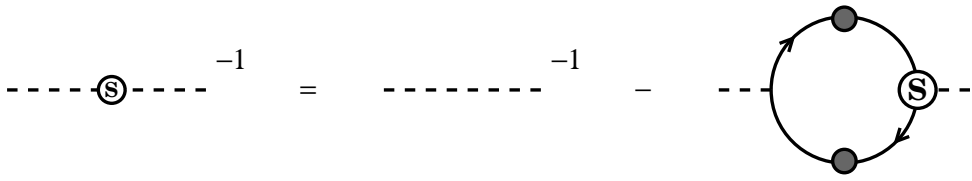}
\caption{The SDE for the scalar propagator $\Delta_{S}(p)$.}
\label{fig_sde_scalar}
\end{figure}

The propagators of the $\sigma$ and $\pi$ fields,  
$\Delta_S$ and $\Delta_P$, are defined, respectively, as follows:
\begin{eqnarray}
\Delta_S(q)&=&-i\int{d}^3x\,{\,e}^{iqx}
\langle 0| T(\sigma(x)\sigma(0))|0\rangle_C,\\
\Delta_P(q)&=&-i\int{d}^3x\,{\,e}^{iqx}
\langle 0| T(\pi(x)\pi(0))|0\rangle_C,
\end{eqnarray}
where the subscript $C$ stands for ``connected.''
The SD  equation for the scalar (pseudoscalar) propagator is given by
\begin{eqnarray}
\Delta_{S(P)}^{-1}(p)=-\frac{N}{G}+\Pi_{S(P)}(p),
\label{delsinv}
\end{eqnarray}
where the (pseudo)scalar vacuum polarization is
\begin{eqnarray}
\Pi_{S(P)}(p)=
i\int^{\Lambda}\frac{d^3k}{(2\pi)^3}\,{\rm Tr}
\left[S(k+p){\Gamma_{S(P)}}(k+p,k)S(k)\Gamma_{0{S(P)}} \right]
\label{sde_scalvac}
\end{eqnarray}
(see Fig.~\ref{fig_sde_scalar}), $S(k)$ is the full fermion
propagator ($S^{-1}(k)=\hat k A(k)-B(k)$), and ${\Gamma_{S(P)}}(k+p,k)$ 
is the fermion-antifermion (Yukawa) vertex (the bare Yukawa vertices are
given by $\Gamma_{0{S}}={\bf 1}$, $\Gamma_{0{P}}=i\gamma_5$, where ${\bf 1}$ is
the identity matrix). 
The absence of kinetic terms for the $\sigma$ and $\pi$ fields in the 
Lagrangian is reflected in the constant bare propagator $-G$. The Yukawa
vertices $\Gamma_S$ and $\Gamma_P$ are defined as the ``fully
amputated'' vertices,
\begin{eqnarray}
S(k)\Gamma_S(k,p)S(p)\Delta_S(k-p)&=&
-\int{d}^3x{d}^3y\,{\,e}^{ikx-ipy}
\langle 0|T(\psi(x)\bar\psi(y)\sigma(0))|0\rangle_C, \\
S(k)\Gamma_P(k,p)S(p)\Delta_P(k-p)&=&-\int{d}^3x{d}^3y\,{\,e}^{ikx-ipy}
\langle 0|T(\psi(x)\bar\psi(y)\pi(0))|0\rangle_C.
\end{eqnarray}
In the symmetric phase of the GNJL model the pseudoscalar and scalar
propagators are degenerate, so are the pseudoscalar vertex and
scalar vertex.

We shall study the SDE for the Yukawa vertex $\Gamma_S$ and scalar
propagator $\Delta_S$ with both the gauge interaction and the
four-fermion interactions treated in the leading order of the $1/N$
expansion. This approximation is obtained by replacing the
Bethe-Salpeter kernel $K$ by planar one photon exchange graph with 
the photon propagator given by Eq.~(\ref{IRphot}) and bare
fermion-photon vertices (see Fig.~\ref{fig:kern}).
In principle the Bethe-Salpeter kernel also contains scalar and
pseudoscalar exchanges. One can question whether such exchanges can
be neglected. In fact, if one includes the ladder like one-scalar
and one-pseudoscalar exchanges in the truncation of the BS kernel
$K$ in the SDE for the Yukawa vertices, 
then such contributions cancel each other exactly in the
symmetric phase. On the other hand, in the equation for the fermion
wave function $A(p)$ these contributions add and must be taken into 
account. Since we take the bare vertex approximation, we need to set
$A(p)=1$ for consistency with the WT identity. In Appendix~\ref{app_nlg} 
we prove the existence of such a nonlocal gauge for the GNJL model in the 
bare vertex approximation and in arbitrary dimensions.\footnote{A version of a 
nonlocal gauge in $D=4$ leading to approximate equality $A=1$ was 
proposed in Ref.\cite{hashimoto}.}
There it is shown also that
four-fermion contributions into the gauge function $\xi$ are
suppressed leading to $\xi(p)=2/3$ (Nash's nonlocal gauge). In what
follows we use the Nash gauge for the photon propagator (\ref{IRphot}).

\begin{figure}
\epsfxsize=9cm
\epsffile[30 400 360 540]{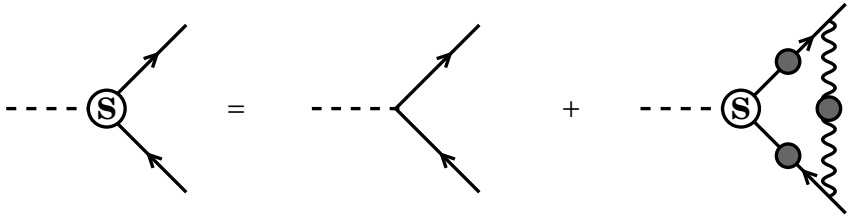}
\caption{The SDE for the scalar Yukawa vertex $\Gamma_S$ in 
the ladder approximation.}
\label{sde_vert}
\end{figure}

The equation for the Yukawa vertex, within the proposed approximation, 
reads
\begin{eqnarray}
{\Gamma_S}(p+q,p)={\bf 1}
+ie^2\int^{\Lambda}\frac{d^3k}{(2\pi)^3}\,\gamma^\lambda S(k+q)
{\Gamma_S}(k+q,k) S(k)
\gamma^\sigma D_{\lambda\sigma}(k-p)
\label{3dsde_ladder}
\end{eqnarray}
(see Fig.~\ref{sde_vert}). In the symmetric phase, the equation for 
the scalar vertex, Eq.~(\ref{3dsde_ladder}), is a self-consistent 
equation if one uses a gauge where the full fermion propagator has the 
form of the free or bare fermion propagator $S(p)=S_0(p)=1/{\hat p}$. 

The invariance under parity and charge conjugation restricts
the form of the Yukawa vertices to the following decomposition
\cite{bu92,gure98}
\begin{eqnarray}
\Gamma_S(p+q,p)&=&
{\bf 1}\left[F_1(p+q,p)
+\left(\hat q\hat p-\hat p\hat q\right)F_2(p+q,p)\right],
\\\label{3dvertfiesdef}
\Gamma_P(p+q,p)&=&
(i\gamma_5)\left[ F_1(p+q,p)
+\left(\hat q\hat p-\hat p\hat q\right)F_2(p+q,p)\right]
\end{eqnarray}
in the symmetric phase.
The two scalar functions $F_i$ are symmetric in the fermion momenta:
\begin{eqnarray}
F_i(p+q,p)\equiv F_i((p+q)^2,p^2,q^2)=F_i(p^2,(p+q)^2,q^2),\qquad i=1,2.
\end{eqnarray}
This is analogous to the four-dimensional case.

Since we are considering the symmetric phase, the $\sigma$ and $\pi$
propagators are identical.
In what follows we neglect the contribution of $F_2$ to the Yukawa vertices.
The validity of this approximation was argued in Ref.~\cite{gure98}
for the four-dimensional case and the analysis 
can be generalized straightforwardly to the three-dimensional case.
Here we only point out that calculating $F_1,F_2$ in $1/N$
perturbation theory reveals that the function $F_1$ contains
logarithmic terms
which build up into the power-law form of the full
solution (see below); on the other hand, $F_2$ does not contain such 
logarithmic terms and thus will not contribute to the leading and 
next-to-leading in $1/N$ order behavior of $\Pi_S$.

Hence, neglecting all functions except $F_1$, we obtain (in Euclidean
formulation) after substituting Eq.~(\ref{IRphot}) with $\xi(p)=2/3$ 
in Eq.~(\ref{3dsde_ladder})
\begin{eqnarray}
F_1(p+q,p)&=&1+\lambda\int\limits_0^{\Lambda}
dk\,\int\frac{d\Omega}{4\pi}\,
\frac{(k^2+q\cdot k)}{(k+q)^2}\frac{1}{|k-p|} F_1(k+q,k), \label{F1eq}
\end{eqnarray}
where $\lambda=32/(3NC\pi^2)$
and where $\int d \Omega$ denotes the usual angular part of the
three-dimensional integration. The equation
for the $\sigma$ boson vacuum polarization is
\begin{eqnarray}
\Pi_S(q)=\frac{2N}{\pi^2}
\int\limits_0^\Lambda dk\int \frac{d\Omega}{4\pi}\,
\frac{(k^2+q\cdot k)}{(k+q)^2} F_1(k+q,k).
\label{vacpolexp1}
\end{eqnarray}

To resolve the angular dependence of the Yukawa vertex function $F_1$ it
is convenient to use an expansion in Legendre polynomials $P_n$
(see also Appendix~\ref{app_twochan}), 
\begin{eqnarray}
F_1(p+q,p)=F_1(p,p+q)&=&\sum_{n=0}^\infty f_n(p,q) P_n(p\cdot q/pq),
\label{firstleg1}
\end{eqnarray}
where in the right-hand side expression $p=\sqrt{p^2}$, $q=\sqrt{q^2}$, 
and $p\cdot q/pq=\cos\alpha$.
Then we follow the arguments of Ref.~\cite{gure98}
and assume that the Yukawa vertex function $F_1(p+q,p)$ depends only
weakly on the angle $p\cdot q/pq$ between fermion and $\sigma$ boson momenta,
so that the set equations for $f_n$ reduces to the equation
for the zeroth-order Legendre coefficient function $f_0$ only.
This is equivalent to
approximating $\Gamma_S$ by its angular average
\begin{eqnarray}
\Gamma_S(p+q,p)=\Gamma_S(p,p+q)\approx
{\bf 1}\int\frac{d\Omega}{4\pi}\,F_1(p+q,p)= {\bf 1} f_0(p,q),
\label{3dcanonic}
\end{eqnarray}
where the function $f_0(p,q)$ depends on the absolute values of 
the vectors $p$, $q$, {\em i.e.}, in it $p=\sqrt{p^2}$, $q=\sqrt{q^2}$.
Accordingly we write 
\begin{eqnarray}
f_0(p,q)=F_{\rm IR}(p,q)\theta(q-p)+F_{\rm UV}(p,q)\theta(p-q),
\label{3dchannelapprox}
\end{eqnarray}
where the functions $F_{\rm IR}$ and $F_{\rm UV}$ satisfy 
integral equations which are given in Appendix~\ref{app_twochan}
(see Eqs.~(\ref{FIRinteq}) and (\ref{FUVinteq})).
Within this approximation, we find that the scalar vacuum polarization 
(\ref{vacpolchanap}) is expressed through the 
function $F_{\rm UV}$ (see Eqs.~(\ref{vacpolchanap}), 
(\ref{a_0}), and (\ref{airauvapprox})):
\begin{eqnarray}
\Pi_S(q)
&=&
\frac{2\Lambda}{\pi^2}\frac{N}{\lambda}\left[F_{\rm UV}(\Lambda,q)-1\right].
\label{3dscalvacpol_eq2}
\end{eqnarray}
The integral Eqs.~(\ref{FIRinteq}) and (\ref{FUVinteq}) can be
reduced to second order differential equations
\begin{eqnarray}
&&p^2 \frac{d^2}{dp^2} F_{\rm IR}+2p \frac{d}{dp}F_{\rm IR}
+\lambda \frac{p^2}{2q^2}F_{\rm IR}=0,
\label{FIRdifeq}
\\
&&p^2 \frac{d^2}{dp^2} F_{\rm UV}+2p \frac{d}{dp}F_{\rm UV}+\lambda
\left(1-\frac{q^2}{2p^2}\right)F_{\rm UV}=0,
\label{FUVdifeq}
\end{eqnarray}
with four boundary conditions.
The infrared and ultraviolet boundary conditions (IRBC and UVBC),
respectively, are
\begin{eqnarray}
\left[p^2\frac{d}{dp}F_{\rm IR}(p,q)\right]_{p=0}=0,
\qquad \left[F_{\rm UV}(p,q)+p\frac{d}{dp}
F_{\rm UV}(p,q)\right]_{p=\Lambda}=1.\label{3dBC}
\end{eqnarray}
There is a continuity and differentiability equation at $p=q$:
\begin{eqnarray}
F_{\rm IR}(q,q)=F_{\rm UV}(q,q),
\qquad
\frac{d}{dp}F_{\rm IR}(p,q)\Biggr|_{p=q}=
\frac{d}{dp}F_{\rm UV}(p,q)\Biggr|_{p=q}.\label{3dcontdiff}
\end{eqnarray}
The equation for $F_{\rm UV}$ can be written as
\begin{eqnarray}
z^2 \frac{d^2}{d z^2} F_{\rm UV}+\left(\lambda-z^2\right)F_{\rm UV}=0,
\qquad z=\sqrt{\frac{\lambda}{2}}\frac{q}{p}. \label{fuv2b}
\end{eqnarray}
The differential Eqs.~(\ref{FIRdifeq}) and (\ref{fuv2b})
and the BC's (\ref{3dBC}) and (\ref{3dcontdiff}) can be solved 
straightforwardly. The solutions are
\begin{eqnarray}
F_{\rm IR}(p,q)&=&Z^{-1}\left(\frac{q}{\Lambda},\omega\right)
\left(\frac{q}{p}\right)
\sin\left(\sqrt{\frac{\lambda}{2}}\frac{p}{q}\right),
\label{3dfir2}
\\
F_{\rm UV}(p,q)&=&
\frac{\pi}{2\sin(\omega\pi/2)}
Z^{-1}\left(\frac{q}{\Lambda},\omega\right)\left(\frac{q}{p}\right)^{1/2}
\nonumber\\
&\times&\left[\rho(\omega)I_{-\omega/2}\left(\sqrt{\frac{\lambda}{2}}
\frac{q}{p}\right)-\rho(-\omega)
I_{\omega/2}\left(\sqrt{\frac{\lambda}{2}} \frac{q}{p}\right)
\right],\label{3dfuv2}
\end{eqnarray}
where $I_{\pm \nu}$ are modified Bessel functions,
and $\omega$ is given by 
\begin{eqnarray}
\omega=\sqrt{1-4\lambda}=\sqrt{1-N_{\rm cr}/N},\qquad
N_{\rm cr}=128/(3C\pi^2).\label{omegadef}
\end{eqnarray}
Furthermore,
\begin{eqnarray}
Z(q/\Lambda,\omega)\equiv
\frac{\pi}{2\sin(\omega\pi/2)}\left[\rho(\omega)
R\left(q/\Lambda,-\omega\right)-\rho(-\omega)
R\left(q/\Lambda,\omega\right)\right], \label{Zdef}
\end{eqnarray}
and
\begin{eqnarray}
\rho(\omega)&\equiv&I_{\omega/2}\left(\sqrt{\frac{\lambda}{2}}\right)
\left[\sqrt{\frac{\lambda}{2}}\cos\sqrt{\frac{\lambda}{2}}
-\frac{1}{2}\sin\sqrt{\frac{\lambda}{2}}\right]
\nonumber\\
&+&I_{\omega/2}^\prime\left(\sqrt{\frac{\lambda}{2}}\right)
\left[\sqrt{\frac{\lambda}{2}}
\sin\sqrt{\frac{\lambda}{2}}\right],\\
R\left(q/\Lambda,\omega\right)&\equiv&\frac{1}{2}\sqrt{\frac{q}{\Lambda}}
\left[
I_{\omega/2}\left(\sqrt{\frac{\lambda}{2}}\frac{q}{\Lambda}\right)
-2\sqrt{\frac{\lambda}{2}}\frac{q}{\Lambda}
I_{\omega/2}^\prime\left(\sqrt{\frac{\lambda}{2}}\frac{q}{\Lambda}\right)
\right].
\end{eqnarray}
The $\sin \omega\pi/2$ results from the Wronskian
between $I_{-\omega/2}(x)$ and $I_{\omega/2}(x)$.

By adopting the approximation (\ref{3dcanonic}) we have obtained
an analytic expression for the Yukawa vertex $\Gamma_S$.
Within this approximation,
the $\sigma$ boson propagator $\Delta_S$ defined by Eq.~(\ref{delsinv})
is related to $\Gamma_S$ via Eq.~(\ref{3dscalvacpol_eq2}).
Such an expression is valid in the symmetric phase of the phase diagram.
\section{Scaling and other properties}\label{sec_scaling}
In the previous section we have obtained nonperturbative solutions
for the Yukawa vertex and scalar propagator within the ladder approximation.
In this section we discuss some important properties of
the Yukawa vertex and scalar propagator.

Let us briefly state our objectives.
First, we apply the Thouless criterion of the symmetry phase
instability in order to derive the critical curve given in 
Eq.~(\ref{critline}). 
Subsequently, we show that near this curve the scalar propagator has a 
scaling form consistent with the general renormalization group 
theory of second order phase transitions.
We find that the anomalous dimension of the propagator of the
composite scalar fields
is $\eta=2-\omega$ with $\omega$ given by Eq.~(\ref{omegadef}).
Moreover, we show that the Yukawa vertex has a scaling form 
consistent with power-law renormalizability.
Second, we derive the peculiar behavior of $\Pi_S$ near $N=N_{\rm cr}$.
The phase transition at $N=N_{\rm cr}$ is known as the CPT and is characterized
by the absence of light unstable modes in the symmetric phase.
Another characteristic feature of the CPT is the scaling law
with essential singularity for the scalar boson and the fermion mass
in the broken phase. This scaling law can be obtained by analytical 
continuation of $\Pi_S$ in $\omega$ across the critical curve at 
$N=N_{\rm cr}$.

In analogy with Ref.~\cite{gure98} we investigate a few specific limits:
\begin{enumerate}
\item[(A)]{
The large flavor limit ($N\rightarrow \infty$), which means that
the gauge interaction is negligible with respect to four-fermion
interactions, {\em i.e.} $\lambda=0$, thus $\omega=1$.}
\item[(B)]{Asymptotic or IR behavior of $\Gamma_S(p+q,p)$ and $\Delta_S(q)$,
{\em i.e.} $p,q\ll \Lambda$.}
\item[(C)]{The behavior of $\Pi_S$ at the critical coupling
$\lambda=\lambda_c=1/4$, thus $\omega=0$.}
\item[(D)]{The behavior of $\Pi_S$
for $\lambda>\lambda_c$, $\omega=i\nu$,
$\nu=\sqrt{4\lambda-1}$, {\em i.e.}, analytic
continuation across the critical curve at $\lambda=\lambda_c$.}
\end{enumerate}
\subsection{Large flavor limit}\label{seclargeN}
In the large flavor limit, the four-fermion interactions
completely govern the dynamical breakdown of ``chiral''
symmetry.
In this limit $\omega=1$ ($\lambda=0$), thus the Yukawa vertex
(\ref{3dsde_ladder}) is $\Gamma_S(p+q,p)=1$.
Consequently, we obtain an expression for $\Pi_S$
from Eq.~(\ref{vacpolchanap}) by using Eq.~(\ref{airauvapprox})
and $F_{\rm UV}(p,q)=F_{\rm IR}(p,q)=1$ at $\omega=1$.
This leads to
\begin{eqnarray}
\Pi_S(q)=\frac{2N\Lambda}{\pi^2}\left[1-\frac{4q}{3\Lambda}
+\frac{q^2}{2\Lambda^2}\right]. \label{vacpolNinfty}
\end{eqnarray}
This expression is obtained by making use of the
approximation (\ref{airauvapprox}).
Naturally, the expression for $\Pi_S(q)$ can be obtained by evaluating
Eq.~(\ref{sde_scalvac}) with $\Gamma_S=1$.
The result is
\begin{eqnarray}
\Pi_S(q)=\frac{2N\Lambda}{\pi^2}\left[1-\frac{\pi^2}{8}\frac{q}{\Lambda}
+\frac{q^2}{3\Lambda^2}\right],\label{exactlead}
\end{eqnarray}
see, {\em e.g.}, Ref.~\cite{hakoko91} and references therein \cite{kiya90}.
Since only the first two terms on the right-hand side of
Eqs.~(\ref{vacpolNinfty}) and (\ref{exactlead})
are important in the IR ($q\ll \Lambda$),
these equations differ about $10\%$.
\subsection{Asymptotic behavior and scaling}\label{subseq:B}
For values $0<\omega <1$,
the asymptotic behavior or IR behavior of $\Gamma_S$ and $\Pi_S$ with
$(q/\Lambda)^\omega \gg q/\Lambda$
can be derived by first considering
the $q\ll \Lambda$ limit of Z, Eq.~(\ref{Zdef}):
\begin{eqnarray}
Z\approx
\frac{\pi}{2\sin(\omega\pi/2)}\left(\frac{q}{\Lambda}\right)^{1/2}
C(\omega)
\sinh \left[\frac{\omega}{2}\ln \frac{\Lambda}{q}+\delta(\omega)\right],
\label{Zapprox}
\end{eqnarray}
where
\begin{eqnarray}
\delta(\omega)&=&\frac{1}{2}
\ln \frac{\rho(\omega)(1+\omega)\Gamma(1+\omega/2)}{
\rho(-\omega)(1-\omega)\Gamma(1-\omega/2)}-\frac{\omega}{4}
\ln \frac{\lambda}{8},
\label{deltadef}\\
C(\omega)&=&
\sqrt{
\frac{\rho(\omega)\rho(-\omega)(1-\omega^2)}{
\Gamma(1+\omega/2)\Gamma(1-\omega/2)}}.\label{Cdef}
\end{eqnarray}
In this limit, the function $F_{\rm UV}(p,q)$
with fermion momentum $p=\Lambda$
can be expressed as
\begin{eqnarray}
F_{\rm UV}(\Lambda,q)\approx \frac{2}{1+\omega}
+\frac{2\omega}{(1-\omega^2)}\left(1-\coth y\right),\qquad
y=\frac{\omega}{2}\ln \frac{\Lambda}{q}+\delta(\omega).
\end{eqnarray}
Thus, by using Eq.~(\ref{3dscalvacpol_eq2}),
the asymptotic form for $\Pi_S$ reads
\begin{eqnarray}
\Pi_S(q)&\approx& \frac{2N\Lambda}{\pi^2}
\left[\frac{4}{(1+\omega)^2}+\frac{8\omega}{(1-\omega^2)^2}
\left(1-\coth y\right)\right].\label{PiSasym1}
\end{eqnarray}
Hence
\begin{eqnarray}
\Pi_S(q)&\approx& \frac{2N\Lambda}{\pi^2}
\left[\frac{1}{g_c}-B(\omega)\left(\frac{q}{\Lambda}\right)^\omega
+{\cal O}\left((q/\Lambda)^{2\omega}\right)
+{\cal O}\left((q/\Lambda)^2\right)\right],\qquad q\ll \Lambda,
\label{vacasym}
\end{eqnarray}
where
\begin{eqnarray}
g_c=\frac{(1+\omega)^2}{4},\quad B(\omega)\equiv \frac{16\omega}
{(1+\omega)^3(1-\omega)}\frac{\rho(-\omega)}{\rho(\omega)}
\frac{\Gamma\left(1-\frac{\omega}{2}\right)}{
\Gamma\left(1+\frac{\omega}{2}\right)}
\left(\sqrt{\frac{\lambda}{8}}\right)^\omega.
\end{eqnarray}
One can show that $B(1)=4/3$, which is in agreement
with Eq.~(\ref{vacpolNinfty}).
The expression (\ref{vacasym}) for the asymptotic behavior of $\Pi_S(q)$ 
is valid for $0<\omega\leq 1$, 
but not for $\omega=0$ ($\lambda=\lambda_c$).

The inverse propagator $\Delta_S^{-1}$ that follows from 
Eqs.~(\ref{delsinv}) and (\ref{vacasym}) is given by
\begin{eqnarray}
\Delta_S^{-1}(q)\approx -\frac{2 B(\omega)N\Lambda}{\pi^2}
\left[\frac{1}{B(\omega)}\left(\frac{1}{g}-\frac{1}{g_c}\right)
+\left(\frac{q}{\Lambda}\right)^{\omega}\right]. 
\label{scalingDelta_S}
\end{eqnarray}
The instability of the symmetric phase is signalized by the vanishing of
$\Delta^{-1}_S(q=0)$. This is nothing else than the Thouless criterion for
a phase transition of the second kind \cite{thouless} which leads to 
the critical curve
\begin{equation}
   g=g_c,\quad 0<\omega<1\quad  (N> N_{\rm cr}),\quad g>\frac{1}{4}.
\end{equation}
Thus the curve $g=g_c$ is a line of UV stable fixed points.
On the critical line the scalar propagator scales as
\begin{equation}
\Delta_S(q)\approx -\frac{\pi^2}{2 B(\omega)N\Lambda}\left(\frac{\Lambda}
{q}\right)^{2-\eta},\qquad \eta=2-\omega,
\label{prop_critline}
\end{equation}
where $\eta$ is the anomalous dimension.

On the other hand, one can see that on the line 
$\omega=0\,(N=N_{\rm cr}),\,
g<1/4$, $\Delta^{-1}_S(q=0)$ does not vanish. Nevertheless, as we shall
show in Sec.~\ref{sec_rescpt}, this line is also the phase transition
line but of a special type.

The scaling form for $\Gamma_S$ is obtained by
considering only the leading term in Eq.~(\ref{Zapprox}).
Thus the $Z$ function scales as
\begin{eqnarray}
Z(q/\Lambda,\omega)\approx
\frac{\pi}{2\sin(\omega\pi/2)}
\frac{\rho(\omega)}{2}\frac{(1+\omega)}{\Gamma(1-\omega/2)}
\left(\frac{\lambda}{8}\right)^{-\omega/4}
\left(\frac{q}{\Lambda}\right)^{(1-\omega)/2}.
\end{eqnarray}
In this way the Yukawa vertex can be written
as
\begin{eqnarray}
\Gamma_S(p+q,p)\approx {\bf 1}\left(\frac{\Lambda}{q}\right)^{(\eta-1)/2}
\left[{\cal F}_{\rm IR}(p/q)\theta(q-p)+{\cal F}_{\rm UV}(q/p)\theta(p-q)
\right],\label{yukawavert_scalform}
\end{eqnarray}
where, for $p,\,q\ll\Lambda$,
\begin{eqnarray}
F_{\rm IR}(p,q)\approx \left(\frac{\Lambda}{q}\right)^{(\eta-1)/2}
{\cal F}_{\rm IR}(p/q),\qquad
F_{\rm UV}(p,q)\approx \left(\frac{\Lambda}{q}\right)^{(\eta-1)/2}
{\cal F}_{\rm UV}(q/p), \label{scalformyukvert}
\end{eqnarray}
and
\begin{eqnarray}
{\cal F}_{\rm IR}(p/q)&=&\frac{2\sin(\omega\pi/2)}{\pi}
\frac{2}{\rho(\omega)}
\frac{\Gamma(1-\omega/2)}{(1+\omega)}
\left(\frac{\lambda}{8}\right)^{\omega/4}
\left(\frac{q}{p}\right)
\sin\left(\sqrt{\frac{\lambda}{2}} \frac{p}{q}\right),\\
{\cal F}_{\rm UV}(q/p)&=&
\frac{2}{\rho(\omega)}
\frac{\Gamma(1-\omega/2)}{(1+\omega)}
\left(\frac{\lambda}{8}\right)^{\omega/4}
\left(\frac{q}{p}\right)^{1/2}
\nonumber\\
&\times & \left[\rho(\omega)I_{-\omega/2}\left(\sqrt{\frac{\lambda}{2}}
\frac{q}{p}\right)-\rho(-\omega)I_{\omega/2}
\left(\sqrt{\frac{\lambda}{2}} \frac{q}{p}\right)\right].
\end{eqnarray}
An important consequence of the scaling behavior of the 
scalar propagator (Eq.~(\ref{prop_critline})) 
and of the Yukawa vertex (Eq.~(\ref{yukawavert_scalform}))
is that, in using them, one finds that the four-fermion 
scattering amplitudes scale as
\begin{eqnarray}
\Gamma_S(p_1+q,p_1)\Delta_S(q)\Gamma_S(p_2,p_2+q)\propto \frac{1}{q},
\qquad p_1,\,p_2\ll q\ll\Lambda.
\end{eqnarray}
This scaling form reveals the long range nature and 
power-law renormalizability of the four-fermion interactions 
at the phase transition line \cite{hakoko93}.
\subsection{At the critical coupling}
At the critical value of $\lambda$, {\em i.e.}, $\omega=0$
($\lambda_c=1/4$),
we can derive in analogy with Ref.~\cite{gure98}
that for $p\gg q$
\begin{eqnarray}
(\omega=0)\qquad
F_{\rm UV}(p,q)\approx
2\left(\frac{p}{\Lambda}\right)^{-1/2}\left[
\frac{\epsilon_3-2+\ln(p/q)}{\epsilon_3-\ln(q/\Lambda)}
+{\cal O}\left(q^2/p^2\ln (q/p)\right)
\right],\label{fuvwzero}
\end{eqnarray}
where
\begin{eqnarray}
\epsilon_1&=&
I_{0}\left(\sqrt{1/8}\right)
\left[\sqrt{1/8}
\cos\sqrt{1/8}-\frac{1}{2}\sin\sqrt{1/8}\right]
+I_{0}^\prime\left(\sqrt{1/8}\right)
\left[\sqrt{1/8}\sin\sqrt{1/8}\right],
\\
\epsilon_2&=&
K_{0}\left(\sqrt{1/8}\right)
\left[\sqrt{1/8}
\cos\sqrt{1/8}-\frac{1}{2}\sin\sqrt{1/8}\right]
+K_{0}^\prime\left(\sqrt{1/8}\right)
\left[\sqrt{1/8}\sin\sqrt{1/8}\right],
\\
\epsilon_3&=&2-\gamma+\frac{5}{2}\ln2-\frac{\epsilon_2}{\epsilon_1},
\end{eqnarray}
with $\gamma$ the Euler gamma
and $K_0$ the modified Bessel function of the third kind.

In the infrared,
{\em i.e.}, $q\ll\Lambda$, $\Pi_S$ can be written as
\begin{eqnarray}
(\omega=0)\qquad
\Pi_S(q)\approx
\frac{2N\Lambda}{\pi^2}\left[4+\frac{16}{\ln(q/\Lambda)-\epsilon_3}
+{\cal O}\left(q^2/\Lambda^2\ln (q/\Lambda)\right)\right]. \label{vacpolwzero}
\end{eqnarray}
This straightforwardly follows from the insertion of Eq.~(\ref{fuvwzero})
in Eq.~(\ref{3dscalvacpol_eq2}).
\subsection{Analytic continuation across the critical curve}
Since the expression for the $\sigma$ boson vacuum polarization
is symmetric under replacement of $\omega$ by $-\omega$,
it can be analytically continued to the values $\lambda> \lambda_c$.
This holds in replacing $\omega$ by $i\nu$ in Eq.~(\ref{3dscalvacpol_eq2})
with $F_{\rm UV}$ given by Eq.~(\ref{3dfuv2}),
where
\begin{eqnarray}
\nu=\sqrt{4\lambda-1}. \label{nudef}
\end{eqnarray}
In the infrared ($q\ll\Lambda$),
it means that
$\Pi_S$ can be written as
\begin{eqnarray}
\Pi_S(q)\approx
\frac{2N\Lambda}{\pi^2}
\left[\frac{4(1-\nu^2)}{(1+\nu^2)^2}-\frac{8\nu}{(1+\nu^2)^2}
\cot y\right],\qquad
y=\frac{\nu}{2}\ln \frac{\Lambda}{q}+\nu \phi(\nu^2),\label{vacpoltach}
\end{eqnarray}
where we have used Eq.~(\ref{PiSasym1}) with $\omega\rightarrow i\nu$,
and where $\phi(\nu^2)=\delta(i\nu)/i\nu$.

The four limits of $\Pi_S$ described above are very useful for 
illustrating the resonance structure of the bound states and peculiar 
dynamics of the CPT, see Sec.~\ref{sec_rescpt}.

To conclude this section
let us mention that at zero $\sigma$ boson momentum ($q=0$),
we obtain
\begin{eqnarray}
\Gamma_S(p,p)=F_{\rm UV}(p,q=0)
=\frac{2}{1+\omega}\left(\frac{p}{\Lambda}\right)^{-(1-\omega)/2}.
\end{eqnarray}

\section{Light resonances and the conformal
phase transition}\label{sec_rescpt}
In this section we analyze the behavior of the $\sigma$ boson propagator
near the critical line in the symmetric phase ($g\leq g_c$),
where the $\sigma$ and $\pi$ boson are degenerate.
We will show that for $\omega>0$ ($N>N_{\rm cr}$) the scalar composites 
($\sigma$ and $\pi$ bosons) are resonances (unstable modes)
described by a complex pole in their respective propagators.
The complex pole in $\Delta_S$ should lie on a second or higher Riemann sheet
({\em i.e.}, not on the first (physical) sheet)
of the complex plane of the Minkowskian momentum $p^2$, because unitarity
(causality) 
demands that $\Delta_S(p)$ is analytic in the upper half of the complex 
$p_0$-plane, where $p_0$ is the ``time'' component of the Minkowski momentum
$p^2=p_0^2-{\vec p}^2$.

From Eq.~(\ref{scalingDelta_S}) the complex pole can be computed.
First we rotate back to Minkowski space,
$p^2\rightarrow p^2_M \exp(-i\pi)$.
Subsequently, the complex poles are given by
\begin{eqnarray}
p^2_M=|m_\sigma|^2 \exp(-i\theta),\qquad \Delta_S^{-1}(p_M)=
-\frac{2N\Lambda}{\pi^2 g}+\Pi_S(p_M)=0.
\end{eqnarray}
The equation for the imaginary part reads
\begin{eqnarray}
0\approx\sin\frac{\omega(\theta+\pi)}{2},
\end{eqnarray}
with the solution
\begin{eqnarray}
\theta\approx-\pi+\frac{2n\pi}{\omega},\label{thetaeq}
\end{eqnarray}
where $n$ is an odd integer. 
Hence for values $0<\omega\leq 1$ it follows from Eq.~(\ref{thetaeq})
that the complex pole does not lie on the physical sheet of $p^2$.\footnote{
We denote the first (physical) Riemann sheet of $p^2$ by angles $\theta$ with
$0\leq -\theta< 2\pi$ (the origin is a branch point with a branch cut
along the positive real axis).}
Since $\cos\omega(\theta+\pi)/2=-1$, we find that the solution for 
$|m_\sigma|$ is
\begin{equation}
\frac{|m_\sigma|}{\Lambda}=\left[\frac{\Delta g}{g_cgB(\omega)}
\right]^{1/\omega},\qquad \Delta g=g_c-g, 
\label{resmsigeq}             
\end{equation}
consequently the critical exponent $\nu=1/\omega$ 
\cite{caclwa91b}.\footnote{The critical exponents $\nu$ and 
$\eta$ coincide with those found in Ref.~\cite{caclwa91b}. 
Note, however, that in Ref.~\cite{caclwa91b} 
$\eta$ was obtained assuming the validity of scaling
relations between the critical exponents of the theory.
Thus the independent computation of $\nu$, $\eta$ 
in the present paper gives, in fact, a proof of the scaling relations.}
Equation (\ref{resmsigeq}) describes how the mass of the pole vanishes
as $g$ is tuned toward the critical line.

The propagator $\Delta_S$ is of the form given by Eq.~(\ref{scalingDelta_S})
and in Minkowski space, with the definition
$p=\sqrt{p^2}$, it can be written as follows:
\begin{eqnarray}
\Delta_S(p)=\frac{\pi^2}{2N\Lambda}
\frac{gg_c}{\Delta g} \left[\frac{-1}{1
+(-1)^{\omega/2}\left(p/|m_\sigma|\right)^\omega}\right],
\end{eqnarray}
with $|m_\sigma|$ given by Eq.~(\ref{resmsigeq}).
Then, the real and imaginary part of $\Delta_S$ are
\begin{eqnarray}
{\rm Re} \left(\Delta_S(p)\right)&=&
-\left(\frac{\pi^2}{2N\Lambda}\frac{gg_c}{\Delta g}\right)
\frac{\left[1+(p/|m_\sigma|)^\omega 
\cos \varphi\right]}{(p/|m_\sigma|)^{2\omega}
+2(p/|m_\sigma|)^\omega \cos\varphi +1},\\
{\rm Im} \left(\Delta_S(p)\right)&=&
-\left(\frac{\pi^2}{2N\Lambda}\frac{gg_c}{\Delta g}\right)
\frac{(p/|m_\sigma|)^\omega \sin \varphi}{(p/|m_\sigma|)^{2\omega}
+2(p/|m_\sigma|)^\omega \cos\varphi +1},\label{impart}
\end{eqnarray}
where $\varphi=\pi\omega /2$.
The absolute value of the imaginary part has a
maximum at $p=|m_\sigma|$ and the maximum is
\begin{eqnarray}
{\rm Im}\left(\Delta_S(|m_\sigma|)\right) =
-\frac{\pi^2}{2N\Lambda} \frac{gg_c}{\Delta g}
\frac{\sin\varphi}{2\left[\cos\varphi+1\right]}. \label{maxp}
\end{eqnarray}
This shows that when $g$ approaches $g_c$ from below ($g\uparrow g_c$),
$|m_\sigma|$ goes to zero ($|m_\sigma|\rightarrow 0$)
and that the maximum of the absolute value of the imaginary part of
$\Delta_S$ approaches infinity
($-{\rm Im}\,\Delta_S(|m_\sigma|) \rightarrow \infty$).

We define a width over mass ratio $\Gamma/|m_\sigma|$
as follows
\begin{eqnarray}
\frac{\Gamma}{|m_\sigma|}=\frac{p_{+}}{|m_\sigma|}-\frac{p_{-}}{|m_\sigma|},
\qquad{\rm Im}\left(\Delta_S(p_{\pm})\right)=\frac{1}{2}
{\rm Im}\left(\Delta_S(|m_\sigma|)\right). \label{widtheq}
\end{eqnarray}
Thus the width is the difference between the momenta at which
$-{\rm Im}(\Delta_S)$ equals $1/2$ of the maximum value of $-{\rm Im}
(\Delta_S)$.
Solving Eq.~(\ref{widtheq}) by making use of Eqs.~(\ref{impart}) and 
(\ref{maxp})
gives
\begin{eqnarray}
\frac{\Gamma}{|m_\sigma|}=
\left[2+\cos\varphi+\sqrt{(2+\cos\varphi)^2-1}\right]^{1/\omega}
-\left[2+\cos\varphi-\sqrt{(2+\cos\varphi)^2-1}\right]^{1/\omega}.
\end{eqnarray}
Thus, as the mass scale of the pole 
is made small by approaching the critical line, the resonance is not 
described by a narrow Breit-Wigner type, 
because the width over mass ratio is rather large. Consequently, the 
resonance does not have the Lorentzian shape which is a characteristic 
feature of the Breit-Wigner resonance (note that even in pure NJL model 
($\omega=1$) the resonance is not narrow in contrast to four-dimensional NJL 
model). The above expression shows also that $\Gamma/|m_\sigma|$
increases when $\omega\rightarrow 0$, and the resonance becomes broader. 

\begin{figure}
\epsfxsize=11cm
\epsffile{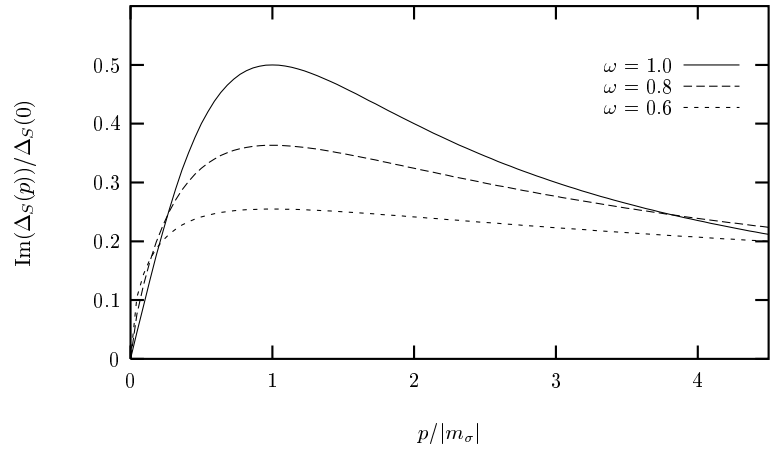}
\caption{The response function ${\rm Im}(\Delta_S(p))/\Delta_S(0)$ vs. 
$p/|m_\sigma|$ for $\omega=1.0$, $\omega=0.8$, $\omega=0.6$.}
\label{res_fig}
\end{figure}

A description of the resonance structure is provided by a plot of 
${\rm Im}\Delta_S(p)$.
This is illustrated in Fig.~\ref{res_fig} in which 
${\rm Im}(\Delta_S(p))/\Delta_S(0)$ is drawn as a function of the energy 
scale $p/|m_\sigma|$ for various values of $\omega$.

\subsection{Absence of light resonances near {\protect\boldmath
$N_{\rm cr}$}}\label{again_sec_cpt3}
The existence of light resonances whose mass vanishes as
the transition is approached from the side of symmetric phase in
(2+1)-dimensional theories is relevant for describing spin excitations
in high-$T_c$ cuprate superconductors (see the paper by Kim and Lee
\cite{kile99} and references therein). Such resonances can be considered 
as precursors of the antiferromagnetic transition.
It is known that QED$_3$ by itself cannot give rise to light excitations
in the symmetric phase \cite{aptewij95,miya97,GMS}.
This is one of the main features of the so-called conformal
phase transition: the absence of light excitations
(composites) in the symmetric phase as the transition is 
approached (in the broken phase massless ``normal'' Goldstone 
bosons appear). This unusual behavior can be attributed to the
long range nature of the gauge interaction in the model under 
consideration.
Another characteristic feature of the CPT is the scaling law with 
essential singularity for the dynamical fermion mass in the broken 
phase \cite{miya97}.

From the side of the symmetric phase there is no sign indicating
the occurrence of a phase transition. This means that the correlation
length remains finite in the symmetric phase even close
to the critical point (the Thouless criterion is not valid).
In QED$_3$ the CPT occurs at $\lambda=\lambda_c$ ($N=N_{\rm cr}$)
where the symmetry is dynamically broken by a ``marginal'' operator
(a long range interaction).
Though continuous, the CPT is not a second order phase transition.
This is reflected by the singular behavior of some of the critical
exponents ({\em e.g.}, $\nu$ and $\beta$, see Ref.~\cite{caclwa91b})
as $\omega$ goes to zero.
The absence of a light complex pole in the $\sigma$ boson
propagator illustrates the CPT in GNJL model in 2+1 dimensions.
At $\omega=0$ the $\sigma$ boson vacuum polarization is given by
Eq.~(\ref{vacpolwzero}) in the infrared.
If there has to be a light excitation in the symmetric phase then there 
must be a complex pole $p_M^2=|m_\sigma|^2 \exp(-i\theta)$
in $\Delta_S$ with $|m_\sigma|\ll \Lambda$ as $g< 1/4$.
From Eq.~(\ref{vacpolwzero}), we then should find zeros of
$\Delta_S^{-1}$ at
\begin{eqnarray}
0&\approx& \left(\frac{1}{g}-4\right)
+\frac{16\left[\ln(\Lambda/|m_\sigma|)+\epsilon_3\right]}{
\left[\ln(\Lambda/|m_\sigma|)+\epsilon_3\right]^2+(\theta+\pi)^2/4},\\
0&\approx&\theta+\pi.
\end{eqnarray}
For $g\leq g_c=1/4$,
there are no solutions satisfying $|m_\sigma|\ll \Lambda$, hence if
there is a pole it will be heavy, {\em i.e.}, $|m_\sigma|\sim \Lambda$.
Therefore at $\lambda=\lambda_c$ and $g<1/4$ there are no light resonances 
in the $2+1$-dimensional GNJL model.

What happens with the $\sigma$-boson propagator if we analytically 
continue it to the values $\lambda>\lambda_c,\,N<N_{\rm cr}$? 
By doing so, we remain in the massless chiral symmetric phase, but
we just end up in the ``wrong vacuum'' 
(the chiral symmetric solution becomes unstable).\footnote{
A phase transition is by definition described by a nonanalytic
behavior in the theory parameters (coupling constants, temperature, etc.). 
Therefore by analytical continuation one cannot go from one phase into
another.}
The $\pi$ and $\sigma$ bosons are tachyons for such a solution. Thus
the border of the stable symmetric solution
$\lambda=\lambda_c,\,g<1/4$ is also the phase transition line.

Let us show that there are indeed tachyons (with imaginary mass $m^2<0$)
when $\lambda>\lambda_c$. For this we need to show that
$\Delta_S$ has a real pole in Euclidean momentum space.
Assuming that the pole lies in the infrared, $|m_\sigma|\ll \Lambda$,
we can use Eq.~(\ref{vacpoltach}),
where $\omega$ has been replaced by $i\nu$, $\nu$ given by
Eq.~(\ref{nudef}).
The tachyonic pole is given by
\begin{eqnarray}
-\frac{2N\Lambda}{\pi^2 g}+\Pi_S(|m_\sigma|)=0.
\end{eqnarray}
From this we derive the solution
\begin{eqnarray}
\frac{|m_\sigma|}{\Lambda}=\exp\left(-\frac{2n\pi}{\nu}-\frac{2\beta}{\nu}
+2\phi(\nu^2)\right),
\end{eqnarray}
where
\begin{eqnarray}
\beta=\tan^{-1} \frac{\nu g}{g-2\lambda(g+\lambda)}.
\end{eqnarray}
As is well established, the tachyon with the largest $|m_\sigma|$
in the physical region corresponds to $n=1$.

It is clear that the tachyons in the symmetric solution appear also when we
cross the upper part ($\lambda<\lambda_c,\,g>1/4$) of the critical curve.
However, the difference between this part of the critical line and
the line $\lambda=\lambda_c,\,g<1/4$ is that we have light composites
(resonances) near the first line while they are absent near the last 
one.

If we now consider the limit $\lambda\downarrow \lambda_c$, {\em i.e.}, 
we approach the transition from the side of the broken phase,
we obtain the scaling law with essential singularity,
\begin{eqnarray}
\frac{|m_\sigma|}{\Lambda}&\approx& \exp\left(\frac{4g}{1/4-g}+2\phi(0)
\right)\exp\left(-\frac{2\pi}{\sqrt{4\lambda-1}} \right).
\end{eqnarray}
This scaling law with essential singularity is obtained
by analytical continuation of the solution in the symmetric phase
($\lambda<\lambda_c$) to the broken phase ($\lambda>\lambda_c$).
Thus the tachyonic (unphysical) solution
in the broken phase leads to a scaling law
which is proportional to the scaling law given by the fermion mass
and $\sigma$ boson mass in the broken phase \cite{Appelquist2}.

\section{Conclusion}\label{conclusion}
In this paper we studied a nonlinear equation for the
running coupling in QED$_3$ which can be considered as the analogue of 
the ladder approximation for the fermion propagator.
We solved our equation both analytically and numerically. 
We found that the vacuum polarization operator, obtained through a 
nonperturbative solution of the equation, has the same infrared asymptotics 
as the one-loop expression: $\Pi(p)\simeq C\alpha/p$, $C\simeq 1+1/14N$. 
Thus, we have showed that a nontrivial IR fixed point persists in the 
nonperturbative solution. Moreover, quantitatively there is only slight
difference between the one-loop result and the nonperturbative
solution even at the number of fermions $N=2$.

We then proceeded with studying the GNJL model in the symmetric phase 
with massless fermions. We solved an equation for the Yukawa vertex in the
approximation where the full Bethe-Salpeter kernel is replaced by the
planar one photon exchange graph with bare fermion-photon vertices. 
The obtained Yukawa vertex was used for calculation 
of the scalar composites propagator. The phase transition curve was
determined from the condition of instability of the symmetric solution.
We established the existence of light excitations 
(resonances) in the symmetric phase for values of $N>N_{\rm cr}\agt 4$ 
($\lambda<\lambda_c$), provided the four-fermion coupling ($g>1/4$) is 
near its critical value along the critical curve (\ref{critline}). 
As $g<1/4$ and $N$ approaches $N_{\rm cr}$ from above the light excitations 
are absent and the situation resembles pure QED$_3$.

The field theoretical models, like QED$_3$ and the GNJL model, often 
appear in the long-wavelength limit of microscopic lattice models used for 
description of high-$T_c$ samples.
For instance, in a spin-charge separation Ansatz for the $t-J$ model, where 
spin is described by fermionic spinons and charge is described by bosonic 
holons (or vice versa), a ``statistical'' U$(1)$ gauge interaction 
appears naturally in the theory along with four-fermion interactions
(see, for example, Refs.\cite{dorey}, \cite{high-T_c}, 
and \cite{kile99}).
It was argued in Ref.~\cite{kile99} 
that QED$_3$, with fermions treated as spinons, might serve as a possible 
candidate for describing the undoped and underdoped cuprates. For physical 
$N(=2)$ the chiral symmetry broken phase of QED$_3$ (with a dynamical mass 
generation) should correspond to an antiferromagnetic ordering in undoped 
cuprates \cite{marston}, 
while the symmetric phase (for larger $N$) would describe some 
kind of a spin liquid. 

Recently spin excitations (particle-hole bound states) have been 
observed in the normal state (and in the 
superconducting state) of underdoped and optimally doped
cuprates such as YBa$_2$Cu$_3$O$_{6+x}$ and La$_{2-x}$Sr$_x$CuO$_4$,
where $x$ is the amount of doping,
see Ref.~\cite{fobosireboivmiakke99} and references therein. 
The dynamic susceptibility $\chi^{\prime\prime}$ 
describing antiferromagnetic correlations near wave vector
$Q=(\pi,\pi)$ has a broad peak whose energy comes down
as the doping is reduced. The height of the peak increases as the doping is
reduced and the antiferromagnetic transition approached.

As was proposed in Ref.~\cite{kile99}, 
QED$_3$ could describe these particle-hole
excitations.
However, from the point of view of the present paper, pure 
QED$_3$ cannot be applied for describing such spin excitations because 
of absence of light resonances in the symmetric phase of the model. 
In our opinion, the 
GNJL model serves this purpose better since light 
excitations appear near the critical curve (\ref{critline}) on both sides. 
Moreover, the mass of resonances decreases as the phase transition is 
approached (along the trajectory $N$, or $\lambda$, is fixed and 
$g\uparrow g_c$) while their peaks become sharper as $g$ approaches $g_c$.
All these features are in qualitative agreement with the experimental 
picture if we assume that the four-fermion coupling $g$ depends on the 
doping in such a way that $g$ increases when the doping is reduced.

A problem, however, is that, in case of cuprate superconductors,
the physically relevant number of flavors equals two ($N=2$)
which is less than $N_{\rm cr}\sim 4$. This means that one would get 
the dynamically broken symmetry phase corresponding to the N\'eel ordered 
state at any doping in both QED$_3$ and the GNJL model.
Kim and Lee \cite{kile99} proposed a mechanism to lower $N_{\rm cr}$ 
(and make $N_{\rm cr}<2$) in pure QED$_3$ by taking into account the 
effect of doping which 
screens the time-component of the gauge field and halves $N_{\rm cr}$, due
to additional coupling of the gauge field to charged scalar fields.
However, another way out of such a dilemma appears if we invoke the arguments
of Appelquist {\em et al.} \cite{Appelquist3} that ladder SD equations 
usually overestimate the critical value $N_{\rm cr}$. These authors suggest 
that the true critical value is $N_{\rm cr}=3/2$. Thus for the physical case 
of $N=2$ the spontaneous symmetry breaking does not occur and the system 
will be in the symmetric phase when the doping exceeds some critical value.
It would be quite interesting to find out a truncated set of SD equations
giving such a small critical $N_{\rm cr}$. 
\acknowledgements{
We would like to acknowledge V.~A.~Miransky for useful and stimulating
discussions and for bringing a paper (Ref.~\cite{Appelquist3}) 
to our attention.
We thank V. de la Incera, V.~A.~Miransky and M.~Winnink for carefully reading 
the manuscript.
V.~P.~G. is grateful to the members of the Department of Physics of
the Nagoya University, especially K.~Yamawaki, for their
hospitality during his stay there.
His research has been supported in part by Deutscher Academischer
Austauschdienst (DAAD) grant, by the National Science Foundation (USA) 
under grant No. PHY-9722059, and by the Grant-in-Aid of Japan Society 
for the Promotion of Science (JSPS) No. 11695030.
He wishes to acknowledge JSPS for financial support.
}
\appendix
\section{Box diagram} \label{app:box}
In this appendix we compute the box function $B$ of Eq.~(\ref{eq:bdef}).
We start by contracting the $\gamma$'s in Eq.~(\ref{boxeq1})
and evaluate the traces.
The result is
\begin{eqnarray}
B(p^2,k^2,p\cdot k)=
i\int_M\frac{d^3r}{(2\pi)^3}\,\left[b_1(p,k,r)+b_2(p,k,r)+b_3(p,k,r)\right],
\end{eqnarray}
where
\begin{eqnarray}
b_1(p,k,r)&=&
\biggr[-16(k\cdot r)^2-4k\cdot r(4k\cdot p -p^2)+4(k^2-4k\cdot p)p\cdot r
-24k\cdot r p\cdot r - 16(p\cdot r)^2 \nonumber \\ &&- 24k\cdot rr^2
+4(k^2 - 3k\cdot p + p^2)r^2 - 24p\cdot rr^2 - 12r^4\biggr]
\nonumber\\ &&\times
\frac{1}{(r+p)^2 (r+p+k)^2 (r+k)^2 r^2},\label{b1eq}
\end{eqnarray}
and
\begin{eqnarray}
b_2(p,k,r)&=&
\biggr[
-4k\cdot rp^2 + 4(2k\cdot p + p^2)p\cdot r + 8(p\cdot r)^2
+4k\cdot rr^2 + 4(2k\cdot p + p^2)r^2 
\nonumber\\&&+ 12p\cdot rr^2+4r^4\biggr]
\frac{1}{r^2 (r+p)^4 (r+p+k)^2},\\
b_3(p,k,r)&=&
\frac{
8k\cdot r p\cdot r
- 4k\cdot p r^2 + 4k\cdot r r^2 + 4p\cdot r r^2 + 4r^4}{r^4 (r+p)^2 (r+k)^2}.
\label{b3eq}
\end{eqnarray}
The traces have been performed with the help of 
{\em FeynCalc} \cite{mebode91}.
Subsequently, we cancel the
$r$ dependence in the numerators of Eqs.~(\ref{b1eq})-(\ref{b3eq}) 
without shifting the integration variable.
In this way
the box function $B$ can expressed as follows (in Euclidean formulation):
\begin{align}
&B(-p^2,-k^2,-p\cdot k)=-2\int_E\frac{d^3r}{(2\pi)^3}\,
\Biggr[
\frac{2}{r^2(r+k)^2}+\frac{1}{r^2(r+k+p)^2}-\frac{4}{(r+p)^2(r+k)^2}
\nonumber \\
&+\frac{1}{(r+p)^4}+\frac{2k^2}{r^2(r+p)^2(r+k)^2}
-\frac{k^2}{(r+p)^4(r+k+p)^2}+\frac{4k^2}{(r+k+p)^2(r+p)^2(r+k)^2}
\nonumber\\
&-\frac{2k^4}{r^2(r+k+p)^2(r+p)^2(r+k)^2}
-\frac{2k\cdot p}{r^2(r+p)^2(r+k)^2}+\frac{2k\cdot p}{r^2(r+p)^2(r+k+p)^2}
\nonumber\\
&-\frac{2k\cdot p}{r^2(r+k+p)^2(r+k)^2}
+\frac{4k\cdot p}{(r+k+p)^2(r+p)^2(r+k)^2}
-\frac{4k^2k\cdot p}{r^2(r+k+p)^2(r+p)^2(r+k)^2}
\nonumber\\
&+\frac{2k\cdot r}{r^4(r+k)^2}
-\frac{4k\cdot r}{r^2(r+p)^2(r+k)^2}+\frac{p^2}{r^4(r+k)^2}
-\frac{p^2}{r^2(r+p)^4}
-\frac{p^2}{r^2(r+k+p)^2(r+k)^2}
\nonumber\\
&+\frac{4p^2}{(r+k+p)^2(r+p)^2(r+k)^2}
+\frac{k^2p^2}{r^2(r+p)^4(r+k+p)^2}
-\frac{5k^2p^2}{r^2(r+k+p)^2(r+p)^2(r+k)^2}
\nonumber\\
&-\frac{4k\cdot pp^2}{r^2(r+k+p)^2(r+p)^2(r+k)^2}
-\frac{2k\cdot rp^2}{r^4(r+p)^2(r+k)^2}
-\frac{p^4}{r^4(r+p)^2(r+k)^2}
\nonumber\\
\intertext{\newpage}
&-\frac{2p^4}{r^2(r+k+p)^2(r+p)^2(r+k)^2}
-\frac{4p\cdot r}{r^2(r+p)^2(r+k)^2}
-\frac{2p\cdot r}{r^2(r+k+p)^2(r+k)^2}
\nonumber\\
&-\frac{2p^2p\cdot r}{r^4(r+p)^2(r+k)^2}
-\frac{2r^2}{(r+k+p)^2(r+p)^2(r+k)^2}\Biggr].
\end{align}
For the explicit calculation of the integral, only a handful of
integrals are involved. These integrals are
\begin{eqnarray}
\int_E\frac{d^3r}{(2\pi)^3}\,
\left[\frac{1}{r^4}-\frac{k^2}{r^4(k+r)^2}\right]&=&0,
\\
\int_E\frac{d^3r}{(2\pi)^3}\,
\left[\frac{k^2p^2}{r^4(k+r)^2(p+r)^2}-\frac{k^2}{r^4(k+r)^2} \right]
&=&\frac{k\cdot p}{8kp|k-p|},\\
\int_E\frac{d^3r}{(2\pi)^3}\,
\frac{1}{r^2(k+r)^2}
&=& \frac{1}{8k},\\
\int_E\frac{d^3r}{(2\pi)^3}\,
\frac{1}{r^2(r+k)^2(r+p)^2}
&=& \frac{1}{8kp|k-p|},\label{triangle}\\
\int_E\frac{d^3r}{(2\pi)^3}\,
\frac{1}{r^2(r+k)^2(r+p)^2(r+k+p)^2}
&=&
\frac{1}{8k\cdot p}
\left[\frac{1}{kp|k-p|}-\frac{1}{kp|k+p|}\right],
\label{last}
\end{eqnarray}
where $|k-p|=\sqrt{(k-p)^2}$.
In the computation of the last integral (\ref{last}), 
we first have rewritten the left-hand side as the sum of four so-called
triangle diagrams (diagrams of the form of Eq.~(\ref{triangle})), 
the subsequent integration is then straightforward.

The final result reads
\begin{eqnarray}
&&B(-p^2,-k^2,-p\cdot k)=\frac{1}{k}+\frac{1}{p}-\frac{k\cdot p}{4kp|k+p|}
+\frac{k\cdot p}{4kp|k-p|}-\frac{(2k^2+kp+2p^2)}{2kp|k+p|}
\nonumber\\
&&-\frac{(2k^2+kp+2p^2)}{2kp|k-p|}-
\frac{(2k^4+5k^2p^2+2p^4)}{4(k\cdot p)kp|k+p|}
+\frac{(2k^4+5k^2p^2+2p^4)}{4(k\cdot p)kp|k-p|}. \label{boxBexpr1}
\end{eqnarray}
If an additional integration over the angle between $p$ and $k$ follows
we can simplify this equation because of the symmetry
$p\cdot k \rightarrow -p\cdot k$. However, if we make use of this symmetry,
we should take the principal value for angular integration,
since the $1/p\cdot k$ singularity no longer explicitly cancels.

At various places, we need the following angular 
integrals:\footnote{The angular measure is $\int d\Omega\,=2\pi
\int_0^\pi d\theta\,\sin\theta$, with $\cos\theta=k\cdot p/kp$.}
\begin{eqnarray}
\int\frac{d\Omega}{4\pi}\,
\frac{1}{|k-p|}&=&\frac{1}{\max(k,p)},\qquad
\int\frac{d\Omega}{4\pi}\,
\frac{k\cdot p}{|k-p|}
=\frac{k^2p^2}{3\,\max(k^3, p^3)},\label{firstangeq}\\
\int\frac{d\Omega}{4\pi}\,
\frac{1}{(k+p)^2}&=&\frac{1}{2kp}\ln\frac{k+p}{|k-p|},
\end{eqnarray}
\begin{eqnarray}
\int\frac{d\Omega}{4\pi}\, \frac{k\cdot p}{(k+p)^2}&=&\frac{1}{2}-
\frac{(k^2+p^2)}{4kp}
\ln \frac{k+p}{|k-p|},\\
\int\frac{d\Omega}{4\pi}\, \frac{(k\cdot p)^2}{(k+p)^2}&=&
-\frac{(k^2+p^2)}{4}\left[1-\frac{(k^2+p^2)}{2kp}
\ln \frac{k+p}{|k-p|}\right],
\end{eqnarray}
and the Cauchy principal value integral
\begin{eqnarray}
\,{\cal P}\int\frac{d\Omega}{4\pi}\,
\frac{1}{k\cdot p |k-p|}
=\frac{1}{kp\sqrt{k^2+p^2}}
\ln\left(\frac{k+p+\sqrt{k^2+p^2}}{|k-p|+\sqrt{k^2+p^2}}\right).
\label{cauchy}
\end{eqnarray}
Note that we use the notation 
$k=\sqrt{k^2}$, $p=\sqrt{p^2}$ for the scalar quantities
in the right-hand side expressions of Eqs.~(\ref{firstangeq})-(\ref{cauchy}).

Here, we compute also a general form of the integral over the kernel 
$K(p,k)$ given by Eq.~(\ref{eq:k}), {\em i.e.},
\begin{eqnarray}
I(\delta) &=& \int_0^\Lambda dk\, \frac{k^\delta}{p^\delta} K(p,k) 
\nonumber \\&=&
\int_0^1 dt\, t^\delta
\Biggr[
-\frac{11 t}{6}
-\frac{1}{t}
+\frac{2t^4+5t^2+2}{2t^2 \sqrt{1+t^2}}
\sinh^{-1} t
\Biggr]
\nonumber \\
&+&
\int_{p/\Lambda}^1 dt\, t^{-1-\delta}
\Biggr[
-\frac{11 t}{6}
-\frac{1}{t}
+\frac
{2t^4+5t^2+2}
{2t^2 \sqrt{1+t^2}}
\sinh^{-1} t
\Biggr]
\nonumber \\
&=&
\int_0^1 dt\, \left(t^\delta +t^{-1-\delta}\right)
\Biggr[
-\frac{11 t}{6}
-\frac{1}{t}
+\frac{2t^4+5t^2+2}{2t^2 \sqrt{1+t^2}}
\sinh^{-1} t
\Biggr]
\nonumber \\
&+& \frac{2}{15} \frac{p}{\Lambda} \lim_{t \rightarrow 0}
 t^{2-\delta}
+ {\cal O}\left(\frac{p^2}{\Lambda^2}\right),
\end{eqnarray}
with $-3 \le  \delta \le 2$. Exact solutions exist when $\delta$ 
is an integer.
In that case, one can make the transformation $t \rightarrow \sinh \ln s 
= (s^2-1)/2s$, after which
the integral can be written as a sum of Spence functions. 
The result is 
\begin{eqnarray}
I(0)=\frac{\pi^2}{4} - \frac{5}{2},\qquad 
I(1)=\frac{\pi^2}{8} - \frac{23}{18},\qquad
I(2)=\frac{\pi^2}{64} - \frac{1}{4}.
\end{eqnarray}
\section{Derivation of the nonlocal gauge in the GNJL model}\label{app_nlg}
In this appendix we derive the nonlocal gauge $\xi$ in the GNJL
model in order to set $A=1$, so that the WTI is satisfied if one
uses the bare vertex approximation. For generality we consider the
case of arbitrary dimensions $D$ and in presence of the mass function
$B$.

We introduce the nonlocal gauge function $\xi(k^2)$ by writing the full
photon propagator in the form
\begin{equation}
e^2 D_{\mu\nu}(k)=-\left(g_{\mu\nu}-\zeta(k^2)\frac{k_\mu
k_\nu}{k^2}\right)\frac{d(k^2)}{k^2},
\end{equation}
where $d(k^2)=e^2/(1+\Pi(k^2))$, $\zeta(k^2)=1-\xi(k^2)$.
The SD equation for the fermion wave function renormalization $A$ is
given by
\begin{eqnarray}
A(p^2)&=&1+\frac{1}{p^2}
\int\frac{d^Dq}{(2\pi)^D}\,\frac{A(q^2)}{q^2A^2(q^2)+B^2(q^2)}\nonumber\\
&\times&
\biggr\{\frac{d(k^2)}{k^2}\left[(D-2) p\cdot q+
\left(p\cdot q-\frac{2(p^2q^2-(p\cdot q)^2)}{k^2}\right)
\zeta(k^2)\right]\nonumber\\
&&-p\cdot q\left[\Delta_S(k^2)+\Delta_P(k^2)\right]\biggr\},
\quad k=p-q.
\end{eqnarray}
Introducing the variables $k^2=x+y-2\sqrt{xy}\cos\theta$, $x=p^2$, $y=q^2$,
and performing the integration over all angles except the angle
$\theta$, we get
\begin{eqnarray}
p^2\left(A(p^2)-1\right)&=&C_D\int\limits_0^{\Lambda^2}dy\,
\frac{y^{(D-2)/2}A(y)}
{yA^2(y)+B^2(y)}\int\limits_0^\pi d\theta\,\sin^{D-2}\theta\nonumber\\
&\times&
\biggr\{
d(k^2)\left[\frac{\sqrt{xy}\cos\theta(D-2+\zeta(k^2))}{k^2}
-2xy\frac{\sin^2\theta}{k^4}\zeta(k^2)\right]\nonumber\\
&&-\sqrt{xy}\cos\theta\left[\Delta_S(k^2)+\Delta_P(k^2)\right]\biggr\},
\end{eqnarray}
where $C_D^{-1}=2^D \pi^{(D+1)/2}\Gamma((D-1)/2)$.
Following the works by Kugo {\em et al.} and 
Kondo {\em et al.} \cite{KugoKondo}
(see also Ref.~\cite{Simmons}) we perform now the $\theta$ integration by
parts in terms containing the first power of $\cos\theta$:
\begin{eqnarray}
p^2\left(A(p^2)-1\right)&=&-\frac{C_D}{D-1}\int\limits_0^{\Lambda^2}dy\,
\frac{y^{(D-2)/2}A(y)(2xy)}{yA^2(y)+B^2(y)}\int\limits_0^\pi
d\theta\,\sin^{D-2}\theta
\nonumber\\
&\times&
\biggr\{
\frac{1}{z^{D-1}}\left[\left(z^{D-2}d(z)\zeta(z)\right)^\prime
-(D-2)z^{D-3}\left(d(z)-zd^\prime(z)\right)\right]\nonumber\\
&&-\left[\Delta_S(z)+\Delta_P(z)\right]^\prime\biggr\},
\label{Afunc:eq}
\end{eqnarray}
where the prime denotes the differentiation with respect to $z=k^2$.

The requirement $A(p^2)=1$ is fulfilled by choosing $\zeta(z)$ such
that the expression in curly brackets in Eq.~(\ref{Afunc:eq}) vanishes.
This gives the first order differential equation for $\zeta(z)$ which
is easy to integrate,
\begin{eqnarray}
\zeta(z)=\frac{D-2}{z^{D-2}d(z)}\int\limits_0^zdt\,t^{D-3}
\left[d(t)-td^\prime(t)\right]+\frac{1}{z^{D-2}d(z)}\int\limits_0^z
dt\,t^{D-1}\left[\Delta_S(z)+\Delta_P(z)\right]^\prime
\end{eqnarray}
(the integration constant was fixed by requiring
$\left[z^{D-2}d(z)\zeta(z)\right]\big|_{z=0}=0$ in order to eliminate the
singularity at $z=0$ in $\zeta(z)$).
The last equation finally leads to the following expression for $\xi(z)$:
\begin{eqnarray}
\xi(z)&=&D-1-\frac{(D-1)(D-2)}{z^{D-2}d(z)}\int\limits_0^zdt\,t^{D-3}d(t)
\nonumber\\&&-\frac{1}{z^{D-2}d(z)}
\int\limits_0^zdt\,t^{D-1}\left[\Delta_S(z)+\Delta_P(z)\right]^\prime.
\label{final-xi}
\end{eqnarray}
For $D=3$ we take
\begin{equation}
d(k)=\frac{e^2}{1+\Pi(k)}\simeq \frac{8}{NC}k,\quad k\ll\alpha,
\end{equation}
with $k=\sqrt{k^2}$,
and assume the following form for scalar propagators in the symmetric
phase and near the critical line (see Eq.~(\ref{prop_critline}))
\begin{equation}
\Delta_S(k)=\Delta_P(k)=
-\frac{a}{\Lambda}\left(\frac{\Lambda^2}{k^2}\right)^\gamma,
\label{sigmapeprop}
\end{equation}
where $a$ is some constant and the power $0<\gamma<1$
(Eq.~(\ref{sigmapeprop}) is verified {\it a posteriori} when solving the
SD equation (\ref{delsinv}) for the scalar propagator). We obtain, from
Eq.~(\ref{final-xi}),
\begin{equation}
\xi(k)=\frac{2}{3}-\frac{NC\gamma a}{4(2-\gamma)}
\left(\frac{\Lambda}{k}\right)^{2\gamma-1}.
\label{xi-D=3}
\end{equation}
In absence of the four-fermion interaction we get the famous nonlocal gauge
$\xi=2/3$ \cite{Nash}. Carena {\em et al.} \cite{caclwa91a} 
have included exchanges by the
bare scalar propagators what corresponds to taking $\gamma=1/2$, $a=4/N$
($C=1$ in their leading order of the $1/N$ approximation for the photon vacuum
polarization). Equation~(\ref{xi-D=3}) then gives $\xi=1/3$ in accordance
with their findings.

Our Eq.~(\ref{scalingDelta_S}) for the scalar propagator gives the exponent
$\gamma=\omega/2$ and contribution due to the exchange of scalars into
$\xi(k)$ becomes suppressed (since $\omega<1$) and we are left with Nash's
nonlocal gauge $\xi=2/3$.
\section{Approximation for the Yukawa vertex}\label{app_twochan}
As was mentioned in Sec.~\ref{sec_GNJL2+1}, in order 
to resolve the angular dependence of the Yukawa vertex function $F_1$,
we expand it, together with the kernels of Eqs.~(\ref{F1eq}) 
and (\ref{vacpolexp1}), 
in Legendre polynomials $P_n$.
We write
\begin{eqnarray}
F_1(p+q,p)&=&\sum_{n=0}^\infty f_n(p,q) P_n(\cos\alpha),\qquad
\frac{1}{|k-p|}=\sum_{n=0}^\infty N_n(k,p) P_n(\cos\beta),
\label{Nndef}
\\
\frac{k^2+q\cdot k}{(k+q)^2}&=&\sum_{n=0}^\infty a_n(k,q) P_n(\cos\gamma),
\label{andef}
\end{eqnarray}
where $\cos\alpha=p\cdot q/pq$, $\cos\beta=p\cdot k/pk$, and
$\cos\gamma=q\cdot k/qk$.
The Legendre polynomials $P_n$ satisfy
\begin{eqnarray}
\int \frac{d\Omega}{4\pi}\,
P_m(\cos\alpha)P_n(\cos\alpha)
=\frac{1}{2}\int\limits_{-1}^1 dx\,P_m(x)P_n(x)= \frac{\delta_{mn}}{2n+1}.
\end{eqnarray}
With the above defined expansions, and by making use of the identity
\begin{eqnarray}
\frac{1}{4\pi}\int_0^\pi d\alpha\,\sin\alpha\int_0^{2\pi}d\theta\,
P_n(\cos\alpha)P_l(\cos\beta)=\frac{\delta_{nl}}{2l+1} P_l(\cos\gamma),
\end{eqnarray}
where $\cos\beta=\cos\alpha\cos\gamma+\sin\alpha\sin\gamma\cos\theta$,
Eq.~(\ref{F1eq}) for the Yukawa vertex can be represented as the set
of equations for harmonics $f_l$:
\begin{eqnarray}
f_l(p,q)&=&\delta_{0l}+\lambda
\int\limits_0^\Lambda dk\, N_l(k,p)
\sum_{m=0}^\infty \sum_{n=0}^\infty C_{lmn} a_m(k,q) f_n(k,q),
\label{fncompeq}
\end{eqnarray}
where
\begin{eqnarray}
C_{lmn}&\equiv& \frac{1}{2}\int\limits_{-1}^1 dx\,
P_l(x)P_m(x)P_n(x)
=
\frac{\left(\frac{1}{2}\right)_{s-l}
\left(\frac{1}{2}\right)_{s-m}
\left(\frac{1}{2}\right)_{s-n} s!}{
\left(s-l\right)!\left(s-m\right)!\left(s-n\right)!
\left(\frac{1}{2}\right)_{s}\left(2s+1\right)},
\end{eqnarray}
where $2s=l+m+n$ and $(a)_k\equiv \Gamma(a+k)/\Gamma(a)$.
The coefficients $C_{lmn}$ are zero unless $l+m+n=2s$ is even and a triangle
with sides $l$, $m$, $n$ exists, {\em i.e.},
$|l-m|\leq n \leq l+m$.\footnote{We thank L.~P.~Kok for pointing
out the paper by Askey {\em et al.}\cite{askora86}.}
Furthermore, Eq.~(\ref{vacpolexp1}) can be written as
\begin{eqnarray}
\Pi_S(q)=\frac{2N}{\pi^2}
\int\limits_0^\Lambda dk\,\sum_{n=0}^\infty \frac{a_n(k,q)f_n(k,q)}{2n+1}.
\label{vacpolcompeq}
\end{eqnarray}

Within the approximation (\ref{3dcanonic}), i.e., keeping the zero-order
term $f_0$ only in right-hand sides of Eqs.~(\ref{fncompeq}) and 
(\ref{vacpolcompeq}), we get 
\begin{eqnarray}
f_0(p,q)&=&1+\lambda \int\limits_0^\Lambda dk\,N_0(k,p)a_0(k,q)
f_0(k,q),\label{f0chanap}\\
\Pi_S(q)&=&\frac{2N}{\pi^2}
\int\limits_0^\Lambda dk\,a_0(k,q)f_0(k,q). \label{vacpolchanap}
\end{eqnarray}
The functions $N_0$ and $a_0$ are straightforwardly obtained from
inverting Eqs.~(\ref{Nndef}) and (\ref{andef}).
This gives
\begin{eqnarray}
N_0(k,p)&=&\frac{\theta(k-p)}{k}+
\frac{\theta(p-k)}{p},\label{N0def}
\end{eqnarray}
and
\begin{eqnarray}
a_0(k,q)&=&\int\frac{d\Omega}{4\pi}\,
\frac{(k^2+q\cdot k)}{(k+q)^2}=a_{\rm IR}(k,q)\theta(q-k)+
a_{\rm UV}(k,q)\theta(k-q),\label{a_0}\\
a_{\rm IR}(k,q)&\equiv&\frac{1}{2}+\frac{(k^2-q^2)}{4qk}
\ln\frac{k+q}{q-k},\qquad
a_{\rm UV}(k,q)\equiv\frac{1}{2}+\frac{(k^2-q^2)}{4qk}
\ln\frac{k+q}{k-q}.
\label{airauv}
\end{eqnarray}
In order to be able to solve the equations for $F_{\rm IR}$ and 
$F_{\rm UV}$ given by Eq.~(\ref{3dchannelapprox}), we approximate
the functions $a_{\rm IR}$ and $a_{\rm UV}$ as follows:
\begin{eqnarray}
a_{\rm IR}(k,q)\approx \frac{k^2}{2q^2},\qquad
a_{\rm UV}(k,q)\approx 1-\frac{q^2}{2k^2},\qquad
a_{\rm IR}(q,q)=a_{\rm UV}(q,q)=\frac{1}{2}.
\label{airauvapprox}
\end{eqnarray}
The validity of this approximation is addressed in Sec.~\ref{seclargeN}.

The lowest order harmonic $f_0$ of the particular Legendre expansion
given in Eq.~(\ref{firstleg1}) is expressed 
in terms of the so-called infrared (IR) function $F_{\rm IR}$
and the ultraviolet (UV) function $F_{\rm UV}$, 
see Eq.~(\ref{3dchannelapprox}).
These functions describe the following asymptotic behavior of
the Yukawa vertex:
\begin{eqnarray}
\lim_{p\gg q}\Gamma_S(p+q,p)&=&{\bf 1} \lim_{p\gg q}
F_{\rm UV}(p,q),\\
\lim_{q\gg p}\Gamma_S(p+q,p)&=&{\bf 1} \lim_{q\gg p}
F_{\rm IR}(p,q).
\end{eqnarray}
The fact that both these asymptotic limits of $\Gamma_S$ are described
by $f_0$ through $F_{\rm IR}$ and $F_{\rm UV}$ 
guarantees the validity of the approximation (\ref{3dcanonic}).
This crucial point is explained in more detail in Ref.~\cite{gure98},
where this approximation is referred to as ``the two-channel approximation.''

Then, by making use of Eqs.~(\ref{3dchannelapprox}), (\ref{N0def}),
and (\ref{airauvapprox}) we get
\begin{eqnarray}
(p<q)\qquad
F_{\rm IR}(p,q)&=&1+\lambda \int\limits_0^p dk\,
\frac{k^2}{2p q^2} F_{\rm IR}(k,q)+\lambda \int\limits_p^q dk\,
\frac{k}{2 q^2}
F_{\rm IR}(k,q)\nonumber\\&+&\lambda \int\limits_q^\Lambda dk\,
\frac{1}{k}\left(1-\frac{q^2}{2k^2}\right) F_{\rm UV}(k,q),
\label{FIRinteq}
\\
(p>q)\qquad
F_{\rm UV}(p,q)&=&1+\lambda \int\limits_0^q dk\,
\frac{k^2}{2p q^2} F_{\rm IR}(k,q)+\lambda \int\limits_q^p dk\,
\frac{1}{p}\left(1-\frac{q^2}{2k^2}\right)F_{\rm UV}(k,q)\nonumber\\
&+&\lambda \int\limits_p^\Lambda dk\,
\frac{1}{k}\left(1-\frac{q^2}{2k^2}\right)F_{\rm UV}(k,q),
\label{FUVinteq}
\end{eqnarray}
and for the scalar vacuum polarization (\ref{vacpolchanap}) we can derive
Eq.~(\ref{3dscalvacpol_eq2}).
The integral Eqs.~(\ref{FIRinteq}) and (\ref{FUVinteq})
are equivalent to the second order differential equations
given in Eqs.~(\ref{FIRdifeq}) and (\ref{FUVdifeq}),
with the four boundary conditions 
(\ref{3dBC}) and (\ref{3dcontdiff}).

\end{document}